\documentclass{pas} 
\usepackage{natbib}
\usepackage{amsmath}
\usepackage{txfonts}

\begin{document}
\lefttitle{$R_{\rm e}$. II. UDGs}
\righttitle{Graham}
\jnlPage{xxx}{xxx}
\jnlDoiYr{2025}
\doival{10.1017/pasa.xxxx.xx}

\articletitt{Research Paper}
\title{$R_{\rm e}$. II. Understanding the IC~3475 galaxy type, including ultra-diffuse
  galaxy, structural scaling relations}

\author{\sn{Alister W.} \gn{Graham}$^1$} 
\affil{$^1$ Centre for Astrophysics and Supercomputing, Swinburne 
University of Technology, Hawthorn, Victoria 3122, Australia.}

\corresp{A.W.\ Graham, Email: AGraham@swin.edu.au}

\citeauth{Graham A.W. (2025) $R_{\rm e}$. II. Understanding IC 3475 galaxy
  type,  including ultra-diffuse galaxy, structural scaling relations {\it Publications of the Astronomical Society of Australia} {\bf 00}, xxx--xxx. https://doi.org/10.1017/pasa.xxxx.xx}

\history{(Received xx xx xxxx; revised xx xx xxxx; accepted xx xx xxxx)}

\begin{abstract}
  
It is explained why relatively gas-poor
ultra-diffuse galaxies (UDGs), a subset of IC~3475 galaxy
types, do not have {\it unexpectedly} large sizes but large sizes that are in
line with expectations from the {\it curved} size-luminosity relation
defined by brighter early-type galaxies (ETGs). 
These UDGs extend the faint end of the
(absolute magnitude, $\mathfrak{M}$)-log(S\'ersic index, $n$) and
$\mathfrak{M}$-(central surface brightness, $\mu_{\rm 0}$) 
relations defined by all ETGs, leading to the large effective half-light radii,
$R_{\rm e}$, in these UDGs. 
It is detailed how the scatter in $\mu_{\rm 0}$, at a given $\mathfrak{M}$, 
relates to variations in the galaxies' values of $n$ and effective surface
brightness, $\mu_{\rm e}$.  These variations map into changes in $R_{\rm e}$
and produce the scatter about the $\mathfrak{M}$-$R_{\rm e}$ relation at fixed
$\mathfrak{M}$. 
Similarly, the scatter in $\mathfrak{M}$, at fixed $\mu_{\rm 0}$ and
$n$, can be mapped into changes in $R_{\rm e}$.
The suggestion that there may be two types of relatively gas-poor UDGs
appears ill-founded, arising from the scatter about the 
$\mathfrak{M}$-$\mu_{\rm 0}$ relation. 
The increased scatter about the faint end of the $\mathfrak{M}$-$R_{\rm e}$
relation and the smaller scatter about $\mathfrak{M}$-(isophotal radii,
$R_{\rm iso}$) relations are explained. 
Artificial and potentially misleading 
size-luminosity relations for UDGs are also addressed.
Finally, expected trends with dynamical mass, and evolutionary pathways towards
relatively gas-rich galaxies, are briefly discussed.
Hopefully, the understanding presented here will prove helpful for
interpreting the many low surface brightness galaxies that the Large Synoptic
Survey Telescope will detect.


\end{abstract}


\maketitle


\section{Introduction}\label{Sec_intro}

\citet{1956AJ.....61...69R} introduced the IC~3475
(I) galaxy type after he analysed the large, low surface brightness
(LSB) dwarf-mass early-type galaxy (ETG) IC~3475. 
Many such I-type galaxies, e.g.\ DDO~28 and DDO~29, were subsequently identified in
the David Dunlap Observatory dwarf galaxy catalogue once distances became available
\citep{1975AnA....44..151F}, and many other IC~3475 types were reported 
\citep[e.g.][]{1983ApJS...53..375R, 1984AJ.....89..919S,
  1986AJ.....91...70V}. 
As noted by \citet{1983ApJS...53..375R}, these large LSB galaxies were designated by the letter
`I', encapsulating both ``IC~3475'' and, to a degree, also their (weakly) 
``Irregular'' structure.\footnote{Use of the letter `I' also 
encapsulated the presence of some Population I stars.} 
\citet{1984AJ.....89..919S} noted that IC~3475 was a bright example of the 
huge [LSB] systems in their catalogue \citep{1985AJ.....90.1681B}. 
In recent years, a 
subset of these large LSB galaxies has been referred to as ultra-diffuse galaxies
\citep[UDGs:][]{2015ApJ...807L...2K, 2015ApJ...798L..45V} if their major-axis effective half-light
radius, $R_{\rm e,maj}$, is larger than $\approx$1.5~kpc and their central $B$- or
$g$-band surface brightness, $\mu_{\rm 0}$, fainter than $\approx 24$ mag arcsec$^{-2}$.

This study demonstrates how IC~3475 type galaxies and thus UDGs 
appear as the natural extension of dwarf 
early-type galaxies (dETGs) in key photometric scaling diagrams, with the
dETGs known to be the extension of ordinary high surface brightness ETGs
\citep[][hereafter Paper~I]{2003AJ....125.2936G, 2019PASA...36...35G}.  UDGs, dETGs, and ordinary ETGs follow the same
log-linear (absolute magnitude, $\mathfrak{M}$)-log(S\'ersic index, $n$) and
$\mathfrak{M}$-(central surface brightness, $\mu_{\rm 0}$) relations.  In
\citet{2023MNRAS.522.3588G}, it was mentioned that the UDGs may also follow
the same curved size-luminosity relation \citep{2006AJ....132.2711G} as 
(dwarf and ordinary) ETGs, extending the sequence to fainter magnitudes
(lower stellar masses) and larger $R_{\rm e}$ sizes. Here, this prediction is shown and quantified.

This connection is crucial for assessing the applicability of suggested UDG
formation scenarios, which have included disorganised star formation in
gas-rich but quiescent systems \citep{1980ApJ...242..517G,
  1983ApJS...53..375R}, or "failed" massive galaxies formed in large dark
matter haloes but quenched early, possibly due to feedback or environmental
effects \citep{1978MNRAS.183..341W, 2006MNRAS.368....2D,
  2015ApJ...798L..45V}. Alternatively, it has been proposed that UDGs may be
low-mass dwarf galaxies that expanded through internal processes such as
bursty star formation and strong stellar feedback \citep{1974MNRAS.169..229L,
  1986ApJ...303...39D, 2017MNRAS.466L...1D}.  Similarly, in dense
environments, such as clusters, tidal interactions and ram-pressure stripping
\citep{1972ApJ...176....1G} might transform initially compact galaxies into
diffuse systems by removing gas and inducing a puffing-up of the stellar
distribution \citep{1996Natur.379..613M, 2001ApJ...547L.123M,
  2006MNRAS.369.1021M, 2015MNRAS.452..937Y, 2019MNRAS.485..382C}. Another
formation channel suggests that UDGs naturally form in high-spin haloes,
leading to extended, LSB galaxies without requiring strong environmental
effects \citep{1980MNRAS.193..189F, 2016MNRAS.459L..51A}.

Paper~I explained how the value of $\mathfrak{M}$
associated with the bend in the size-magnitude relation depends on the arbitrary percentage
of light enclosed within the scale radius, which is invariably set at 50 per
cent to give $R_{\rm e}$.  Consequently, this bend does not signify a physical
division or unique formation processes operating above and below this
magnitude, and care needs to be taken when interpreting slopes and bends in
curved size-luminosity diagrams. 
Additional care is required when the sample selection is based on the
galaxies' central
surface brightness, as this can slice the data in such a way that it is
distributed nearly orthogonal to the underlying size-magnitude
trend defined by the population at 
large, giving rise to some misleading results. This is addressed within. 

Just as there is a continuity in structure between dETGs and ordinary ETGs, no
distinct separation exists between IC~3475 type galaxies and UDGs and (dwarf
and ordinary) ETGs in the $\mathfrak{M}$-$R_{\rm e}$ diagram.
Here, UDGs are presented as the natural extension of the early-type galaxy (ETG) sequence to lower luminosities and surface brightnesses. The curved size–luminosity relation, along with consistent trends in Sérsic index and surface brightness, suggests continuity rather than a new class of galaxy.
This study
reveals the reasons behind the increased scatter observed in the (size,
$R_{\rm e}$)-luminosity relation at faint magnitudes.  Furthermore, given
that IC~3475 type galaxies have long been known to display a different
behaviour --- having smaller radii and less scatter --- in diagrams using isophotal radii
\citet[e.g.][their Figures~1 and 2, with $-20 < \mathfrak{M}_B <
  -12$~mag]{1991AnA...252...27B}, the $\mathfrak{M}$-$R_{\rm iso}$ relation 
and its smaller leval of scatter is also presented and explained.  Finally, 
the scatter about the above key scaling relations is used to explain trends in
dynamical mass for dwarf galaxies, including UDGs.

The data used for this paper are described in Section~\ref{Sec_data}, and
the analysis is performed in Section~\ref{Sec_anal}, where the key results are
presented. A review is provided in Section~\ref{Sec_Disc}, along with a
select discussion of some immediate and important implications and a
recognition that the ETG sequence is one of mergers, from primaeval 
galaxies to wet-merger-built dust-rich S0 galaxies and dry-merger-built E galaxies.

\section{Data}\label{Sec_data}

Two compilations of galaxy structural parameters have been used, one for
(dwarf and ordinary) ETGs \citep{2003AJ....125.2936G} and one for UDGs
\citep{2025MNRAS.536.2536B}.

The ETG sample comprises the alleged (dwarf and ordinary) elliptical galaxies from 
\citet{1993MNRAS.265.1013C} and \citet{1994MNRAS.271..523D}, \citet[][no S\'ersic
  indices]{1997AJ....114.1771F}, \citet[][]{1998AnA...333...17B},
\citet[][$R_{\rm e}$ derived from the other S\'ersic
  parameters]{2001AJ....121.1385S}, and the Coma cluster dwarf elliptical
galaxies modelled in 
\citet{2003AJ....125.2936G}.  The parameters displayed in this work pertain to
the $B$-band on the Vega magnitude system and include the galaxies' absolute
magnitude, $\mathfrak{M}_B$, and the central surface brightness, $\mu_{\rm 0,B}$.
For simplicity, no subscript $B$ is assigned to the S\'ersic index $n$ or the
effective half-light radii, $R_{\rm e}$, although, due to radial colour
gradients, these parameters are slightly dependent on the filter used
\citep[e.g.][]{2012MNRAS.421.1007K, 2013MNRAS.430..330H, 2016MNRAS.460.3458K,
  2016AnA...593A..84K}.

The UDG sample consists of 36 `nearly UDGs' (NUDGes)\footnote{This term was
coined by \citet{2024MNRAS.528..608F} and snares galaxies often also considered by
others \citep[e.g.][]{2020ApJ...899...69L, 2022A&A...662A..43V, 2023ApJ...955....1L}.}  presented in
\citet{2025MNRAS.536.2536B}, see also \citet{2024AnA...690A.339M}, plus their
remodelled data for another 28 UDGs previously studied by
\citet{2022MNRAS.517.2231B}, along with data for another 59 UDGs taken from
\citet{2024MNRAS.529.3210B}, see also \citet{2021AnA...654A.105M}.  The NUDGes 
have slightly smaller sizes and slightly brighter
central surface brightnesses than UDGs
and are regarded as mainly dE galaxies \citep{2025MNRAS.536.2536B}. 
This UDG and NUDG data compilation spans stellar 
masses of $7 \lesssim \log(M_{\star}/M_\odot) \lesssim 8.5$ dex, and their
$g$-band (AB) data is used here.
These recent works are selected because they not only propogate the idea that
UDGs should perhaps be separated from (dwarf and ordinary) ETGs --- based on
their sizes ($R_{\rm e,maj}=1.5$~kpc) and central surface brightnesses
$\mu_{\rm 0,g}=24$ mag arcsec$^{-2}$) --- but because they additionally
suggest that there might be two classes of UDGs.  While the former suggestion
is addressed here by reviewing the extension of the size-luminosity relation
to fainter luminosities, the latter development is addressed here by
considering the scatter about the size-luminosity relation.

Given that the UDG $(g-z)_{\rm AB}$ colours in the above works peak at
$\approx 0.9$--1 mag, this data is expected to have $(g-i)_{\rm AB} \approx
0.8\pm0.1$ and $(B-V)_{\rm Vega} \approx 0.6\pm0.1$.
The $g$-band (AB)
magnitudes are transformed into $B$-band (Vega) magnitudes using $B_{\rm AB}=
g_{\rm AB} + 0.35$ \citep{2005AJ....130..873J, 2006A&A...460..339J}, which
should vary by only $\pm0.04$~mag due to the colour spread in the data, and
then using $B_{\rm Vega} = B_{\rm AB}+0.12$ \citep[][their
  Table~8]{1996AJ....111.1748F}.
The UDGs and NUDGes were separated into one of two classes (A or B) by
\citet[][see their section~4.3]{2025MNRAS.536.2536B}, based on an array of
galaxy and associated globular cluster system properties.\footnote{For the
sample from \citet{2024MNRAS.529.3210B}, the class adopted here is that
derived without the number of globular clusters used as a criterion; this was
done as it resulted in a greater number of assigned classes.
This does not imply an irrelevance of the globular cluster systems of UDGs
\citet{2017ApJ...844L..11V, 2018MNRAS.475.4235A, 2020ApJ...899...69L,
  2024AnA...690A.339M}.}
Key factors in separating galaxies into Class A and Class B turned out to be
the offset from the classical dwarf galaxy mass–metallicity relation
\citep[e.g.][]{2013ApJ...779..102K}, the number of associated globular clusters, and the
axial ratio. Class A systems broadly resemble traditional dwarf ETGs, with
structural and chemical properties consistent with formation through internal
processes like supernova-driven feedback \citep{1986ApJ...303...39D}, supporting a
“puffy dwarf” origin \citep{2017MNRAS.466L...1D}. In contrast, Class B galaxies
show signs of early quenching of stars and high globular cluster richness, suggestive of more massive
haloes that failed to form substantial stellar populations, consistent with a
failed galaxy scenario \citep{2015ApJ...798L..45V}. Just five of the three dozen
NUDGes were assigned to Class B.

As with the ETG sample, the most distant UDGs are located in the Coma cluster,
which is $\sim$100 Mpc distant.  Given this proximity, the cosmological
corrections\footnote{These are the $K$-correction, evolutionary correction,
and $(1+z)^4$ surface brightness dimming.} are minor, and only the $(1+z)^4$
surface brightness dimming correction is applied.
Some dETGs, particularly in the field or low-density environments,
can harbor gas and display recent or ongoing star formation
\citep[e.g.][]{2009MNRAS.394.1229C, 2009A&A...498..407G, 2009MNRAS.396.2133K,
  2010MNRAS.409..500O, 2023MNRAS.520.5521S}. 
Field and isolated UDGs often retain significant H{\footnotesize I}
gas reservoirs and show evidence of ongoing or recent star formation, though at low efficiencies
\citep{2022ApJ...941...11K, 2024ApJ...975...91K}. This is not surprising given
they often have an irregular or knotty structure. 
These systems likely represent the actively evolving, gas-retaining
counterparts of the quiescent Class~A UDGs. In this context, Class~A would
span a continuum from still-forming, diffuse dwarfs in the field to faded,
quiescent analogues in higher-density environments.

\begin{figure*}[ht]
\begin{center}
\includegraphics[angle=0, width=1.0\textwidth]{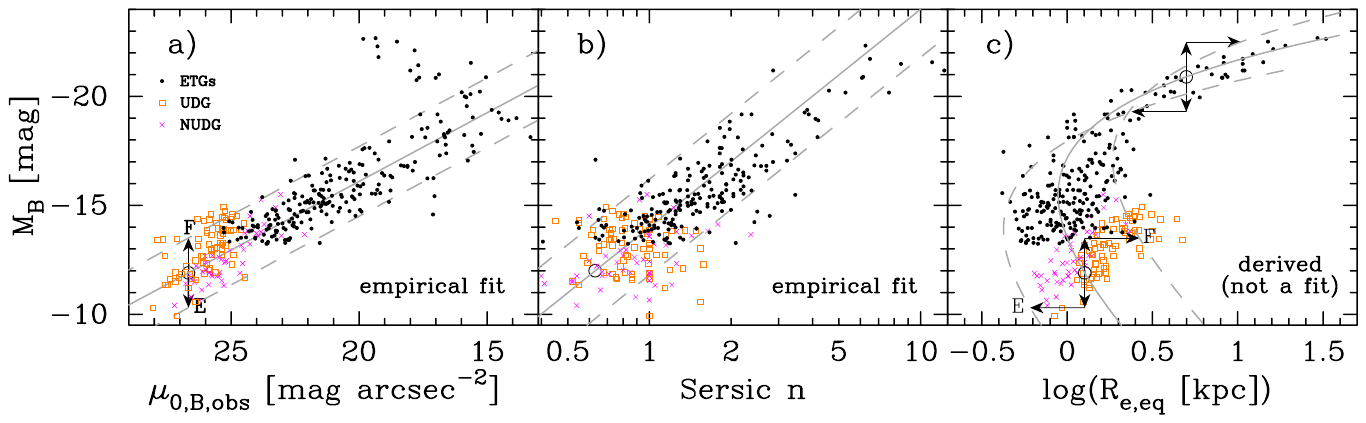}
\caption{Absolute magnitude versus central surface brightness
  (Equation~\ref{Eq_Mmu}, panel a),  S\'ersic index (Equation~\ref{Eq_Mn},
  panel b),
and (equivalent axis) effective half-light radius (Equation~\ref{Eq_MR}, panel
c).
The dashed lines in panel a) ensnare roughly the $\pm2\sigma$ scatter about
the ETGs, and they have been mapped into panel c). The arrows reveal how a
change in absolute magnitude at fixed $\mu_{\rm 0,B}$ and $n$ results in a
corresponding change in $R_{\rm e}$, simultaneously explaining the low and
high levels of scatter at bright and faint absolute magnitudes, respectively.
For the curves in panel c), the central surface brightness (and S\'ersic
index) associated with points E and F (and the lower circle) are the same, but
$R_{\rm e}$ is different because the magnitude is different (see
Equation~\ref{Eq_MR}).  Here, galaxies from \citet{2024MNRAS.529.3210B} and
\citet{2025MNRAS.536.2536B} are designated UDGs if they have $R_{\rm
    e,maj} \ge 1.5$~kpc and $\mu_{\rm 0,g} \ge 24$ mag arcsec$^{-2}$,
  otherwise they are designated NUDGes. The ETG sample compiled by
  \citet[][see Section~\ref{Sec_data}]{2003AJ....125.2936G} contains a few
  (6--9) galaxies that would likely be considered UDGs using this definition.}
\label{Fig1}
\end{center}
\end{figure*}

\section{Analysis and Relations}\label{Sec_anal}

The data points in the left and central panel of Figure~\ref{Fig1} reproduce
the $B$-band (absolute magnitude, $\mathfrak{M}_B$)--(central surface
brightness, $\mu_{\rm 0,B}$) and $\mathfrak{M}_B$--(S\'ersic index, $n$) diagrams
from \citet[][their Figure~9]{2003AJ....125.2936G}. 
The two empirical relations shown here are 
\begin{equation}
\mathfrak{M}_B = 0.63\mu_{\rm 0,B,obs} - 28.7, \sigma=0.8\, {\rm mag} \label{Eq_Mmu}
\end{equation}
and
\begin{equation}
  \mathfrak{M}_B = -10.0\log(n) - 14.0, \label{Eq_Mn} 
\end{equation}
where $\sigma$ denotes the 1-sigma scatter in the $\mathfrak{M}_B$-direction about the
$\mathfrak{M}_B$-$\mu_{\rm 0,B}$ relation. 
These slopes and intercepts
include a slight (5 per cent) variation on the values reported
by \citet{2003AJ....125.2936G}, yielding a slightly more symmetrical scatter
of the ETG data about these lines. 
This will, nonetheless, soon be revisited once the IC~3475 galaxy types are added.
The subscript `obs' has been added to $\mu_{\rm 0,B}$ to differentiate this
observed surface brightness from an upcoming adjusted value.
Galaxies from
from \citet{2024MNRAS.529.3210B, 2025MNRAS.536.2536B} with 
$R_{\rm e,maj} \ge 1.5$~kpc) and central surface brightnesses $\mu_{\rm                                      
  0,g} \ge 24$ mag arcsec$^{-2}$ are designated as UDGs in Figure~\ref{Fig1},
while the remaining galaxies from these two works are denoted NUDGes.

The curved line in Figure~\ref{Fig1}c is not a fit to the data; it is the
solution for \citet{1968adga.book.....S} $R^{1/n}$ light profiles
\citep{1989woga.conf..208C, 2005PASA...22..118G}  
based on Equations~\ref{Eq_Mmu} and \ref{Eq_Mn}.
This curve pertains to the radius equivalent to the geometric-mean of the minor 
and major axis, $R_{\rm e,eq} = \sqrt{R_{\rm e,maj} R_{\rm e,min}}$, and stems from the analytical expression
\begin{equation}
\mathfrak{M} = \langle \mu \rangle_{\rm e,abs} - 2.5\log(2\pi R_{\rm e,eq}^2)- 36.57, \label{Eq_step} 
\end{equation}
with $\langle \mu \rangle_{\rm e,abs}$ the absolute mean effective surface
brightness in units of mag arcsec$^{-2}$ and $R_{\rm e,eq}$ in kpc 
\citep[see][their Equation~12]{2005PASA...22..118G}.
As such, 
$\langle \mu \rangle_{\rm e,abs} = \langle \mu \rangle_{\rm e} -10\log(1+z)$. 
Equation~\ref{Eq_step}  can be re-expressed as 
\begin{equation}
\log R_{\rm e,eq} {\rm (kpc)} = (\langle \mu \rangle_{\rm e,B,abs} - \mathfrak{M}_B)/5.0-7.713 \label{Eq_MR}.
\end{equation}
The difference between $\langle \mu \rangle_{\rm e}$ and $\mu_{\rm 0}$
(and $\langle \mu \rangle_{\rm e,abs}$ and $\mu_{\rm 0,abs}$) 
is a (mathematical, not empirical) function of the S\'ersic model, given by 
\begin{equation}
\mu (R)=\mu_{\rm e}+\frac{2.5b_n}{\ln(10)}\left[\left(R/R_{\rm
    e}\right)^{1/n}-1\right], 
\label{EqSB}
\end{equation}
and, in particular, the S\'ersic index $n$, such that 
\begin{equation}
\langle \mu \rangle_{\rm e} = \mu_{\rm 0} + 2.5b/\ln(10) - 2.5 \log[f(n)], \label{Eq_mumu}
\end{equation}
with the effective surface brightness at $R_{\rm e}$ denoted $\mu_{\rm e}$ and 
\begin{equation}
  \mu_{\rm 0} + 2.5b/\ln(10) = \mu_{\rm e} \label{Eq_meow}
\end{equation}
\citep[see Equations 7, 8 and 9 from][]{2005PASA...22..118G}.
One additionally has that 
\begin{equation}
f(n) = \frac{n {\rm e}^b \Gamma(2n)}{b^{2n}}, 
\end{equation}
with $b$ a function of $n$ obtained by solving 
$\gamma(2n,b) / \Gamma(2n) = 1/2$, and $\Gamma$ is the complete gamma
function. 

Therefore, knowing how $n$ and $\mu_{\rm 0,B}$ vary with $\mathfrak{M}_B$ 
(Equations~\ref{Eq_Mmu} and 
\ref{Eq_Mn}), one knows how $\langle \mu \rangle_{\rm e,B}$ (and $\mu_{\rm
  e}$) varies with $\mathfrak{M}_B$, and hence, 
$R_{\rm e,eq}$ can be derived as a function of $\mathfrak{M}_B$.  This derivation is
shown by the central, solid curve in 
Figure~\ref{Fig1}c, explaining the trend in the data toward 
larger sizes at lower luminosities. Although this curved size-luminosity
relation was presented in \citet{2006AJ....132.2711G}  and
\citet{2008MNRAS.388.1708G}, it is extended here to fainter magnitudes 
to show better the expected increase in size of fainter ETGs/UDGs.
In this panel, and all others bearing the inset text ``derived (not a fit)'',
the curves are not a fit to the data but are derived from empirical fits to
other galaxy parameters coupled with the S\'ersic model. 

The dashed lines in Figure~\ref{Fig1}a roughly capture the $\pm2\sigma$
scatter. 
These lines have been combined with Equation~\ref{Eq_Mn} to produce the dashed
curves in Figure~\ref{Fig1}c. They are insightful because they show how, for a
fixed $n$ and $\mu_{\rm 0,B}$, changes in luminosity are met by changes in 
size.  This 
combination results in shifts roughly along the
size-luminosity relation for galaxies with high S\'ersic indices but perpendicular to it 
for galaxies with low S\'ersic indices. 
At faint magnitudes, where the S\'ersic index is low,
a 2$\sigma$ change in magnitude of 1.6 mag (or roughly a factor of 4 in
luminosity) requires a roughly factor of 2 change in $R_{\rm e,eq}$. 
This explains the broadening of the size-luminosity relation for ETGs at 
low masses and is shown by the labels E and F in Figure~\ref{Fig1}.

\begin{figure*}[ht]
\begin{center}
\includegraphics[angle=0, width=1.0\textwidth]{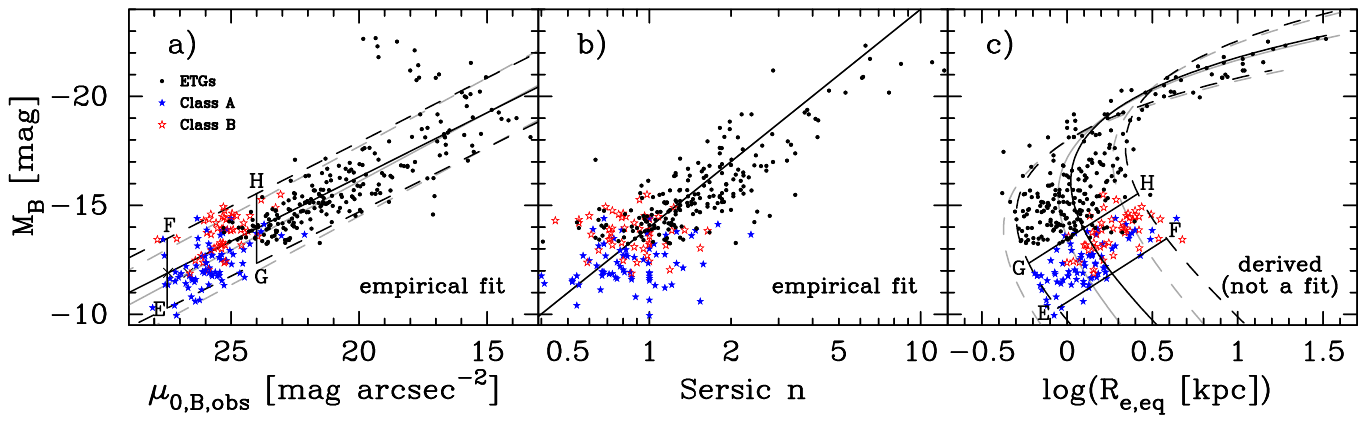}
\caption{Extension of Figure~\ref{Fig1} with the addition of UDG and NUDG data from
  \citet{2025MNRAS.536.2536B} and \citet{2024MNRAS.529.3210B}, and a slightly
  revised $\mathfrak{M}_B$-$\mu_{\rm 0,B,obs}$
  relation (Equation~\ref{Eq_M-mu2}) and thus a slightly revised $\mathfrak{M}_B$-$R_{\rm
    e,eq}$ relation (Equation~\ref{Eq_Mag_R}). The colour-coding tracks the
  two classes (A=blue filled star, B=red open star) 
  assigned to the UDGs and `NUDGes' by \citet{2025MNRAS.536.2536B}.
  The two vertical lines of constant central surface brightness shown in the
  lower left of panel a) map into the diagonal lines seen in panel c), with the
  arrows in Figure~\ref{Fig1}c showing how this occurs. This reveals and
  explains the creation of the misleading size-luminosity relation for UDGs,
  relative to the ETG population at large.
}
\label{Fig2}
\end{center}
\end{figure*}

Figure~\ref{Fig2} introduces the UDGs, which appear to be the extension of 
(ordinary and dwarf) ETGs to fainter magnitudes.  This is most immediately obvious in the
$\mathfrak{M}_B$--$\mu_{\rm 0,B}$ and $\mathfrak{M}_B$--$\log(n)$ diagrams due to the
roughly log-linear behaviour of these relations before the onset of core
depletion in the massive elliptical galaxies built from gas-poor major mergers.
The colour-coding assigned to the UDGs and `NUDGes'
shows the two classes (A=blue filled star, B=red open star) assigned by \citet{2025MNRAS.536.2536B}.

The following slightly revised version of Equation~\ref{Eq_Mmu} better matches the
fuller ensemble of data shown in Figure~\ref{Fig2}a: 
\begin{equation}
\mathfrak{M}_B = 0.59\mu_{\rm 0,B,obs} - 28.1, \sigma=0.8 {\rm mag}. \label{Eq_M-mu2}
\end{equation}
By feeding this empirical relation into Equations~\ref{Eq_MR} and \ref{Eq_mumu},
one derives the following relation shown by the black curve in Figure~\ref{Fig2}c.
\begin{equation} 
\log R_{\rm e,eq} {\rm (kpc)} = 0.139\mathfrak{M}_B - 0.5\log[f(n)] + 0.217b + 1.812, 
\label{Eq_Mag_R}
\end{equation}
with $n=10^{-(14.0+\mathfrak{M}_B)/10.0}$ coming from Equation~\ref{Eq_Mn}.  

\begin{figure*}[ht]
\begin{center}
\includegraphics[angle=0, width=1.0\textwidth]{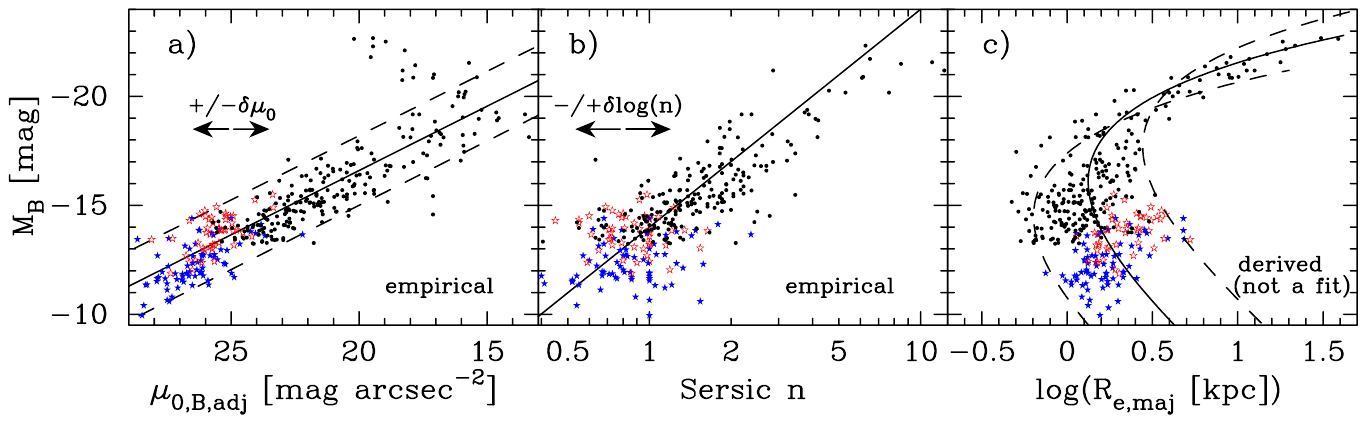}
\caption{Variant of Figure~\ref{Fig2} showing
  $\mu_{\rm 0,B,adj}$ (see Equations~\ref{EqInc} and \ref{Eq_adj}) in panel a)
and 
  $R_{\rm e,maj}$ (see Equation~\ref{Eq_Mag_Rmajor}) in panel c). 
Symbols have the same meaning as in Fig~\ref{Fig2}. 
At fixed $\mathfrak{M}_B$, the (horizontal) scatter $\delta\mu_{\rm 0,B}$ is directly
related to the scatter $\delta\log(n)$ (panel b) and $\delta\mu_{\rm e,B}$ (see Figure~\ref{Fig4}).
}
\label{Fig3}
\end{center}
\end{figure*}

It is noted that a sample collection gap is evident in Figure~\ref{Fig2}a at
$\mathfrak{M}_B>-12$~mag and $\mu_{\rm 0,B} < 24$ mag~arcsec$^{-2}$, leading
to a small triangular cutout that can be seen to propagate into
Figure~\ref{Fig2}c (and Figure~\ref{Fig3}c).  This creates the illusion of a
trend for the UDGs and NUDGes in the $\mathfrak{M}_B$-$R_{\rm e}$ diagram that
is roughly perpendicular to the actual trend for UDGs/NUDGes and (dwarf and
ordinary) ETGs.  That is, Figure~\ref{Fig2}c reveals how the arbitrary sample
selection of the UDGs yields an artificial and misleading size-luminosity
relation.

\subsection{Major axis radii}\label{Sec_Maj}

If one uses the major axis radius, $R_{\rm e,maj}$ ($= R_{\rm e,eq}
\sqrt{b/a}$, where $b/a$ is the minor-to-major axis ratio), 
in the size-luminosity diagram, then an adjustment to Figure~\ref{Fig2}a is
required.
Due to an increased size, a fainter surface brightness is required
to balance the derivation of the magnitude (Equation~\ref{Eq_step}). The
adjusted surface brightness affects $\langle \mu \rangle_{\rm e}$ and
$\mu_{\rm 0}$
equally and is given by
\begin{equation}
  \mu_{\rm 0,B,adj} =  \mu_{\rm 0,B,obs} - 2.5\log(b/a). \label{EqInc}
\end{equation}
This is similar to a correction for disc galaxies where
$b/a \approx cos(i)$ and $i$ is the inclination of the 
disc such that $i=0$ denotes alignment of the galaxy's polar axis with
our line of sight, i.e.\ face-on.  
Although rotation and the presence of faint spiral patterns and bars in some
dwarf galaxies reveal their disc-like nature, most are thought to be triaxial,
and the brighter dETGs with regular isophotes are less triaxial and slightly more spherical 
\citep[e.g.][]{1998ApJ...505..199S}. Basically, 
the rebalancing performed through Equation~\ref{Eq_step} accounts for the larger radii
($R_{\rm e,maj} > R_{\rm e,eq}$) encapturing a fainter `mean effective surface
brightness', which is implemented here by making the central
surface brightness fainter via Equation~\ref{EqInc}.

Observed axis ratios for the dETG sample in \citet{1998AnA...333...17B} have
come from the values reported by \citet{1993AnAS...98..297B}, while for the
remaining UDG/NUDG and (dwarf and ordinary) ETG samples, they have come from
the reported ratios or ellipticities in their respective papers.  The result is
shown in Figure~\ref{Fig3}a, along with the slightly revised line
\begin{equation}
\mathfrak{M}_B = 0.59\mu_{\rm 0,B,adj} - 28.4, \sigma=0.8\, {\rm mag}, \label{Eq_adj}
\end{equation}
and thus a new expression for the $\mathfrak{M}_B$-$R_{\rm e,maj}$ curve 
(Figure~\ref{Fig3}c) that differs slightly from Equation~\ref{Eq_Mag_R} shown in
Figure~\ref{Fig2}c.  This new equation for the major axis $R_{\rm e}$ is given by 
\begin{equation}
\log R_{\rm e,maj} {\rm (kpc)} = 0.139\mathfrak{M}_B - 0.5\log[f(n)] + 0.217b +
1.914. 
\label{Eq_Mag_Rmajor}
\end{equation}

\subsection{Min(d)ing the scatter}\label{Sec_mind}

\begin{figure}[ht]
\begin{center}
\includegraphics[angle=0, width=1.0\columnwidth]{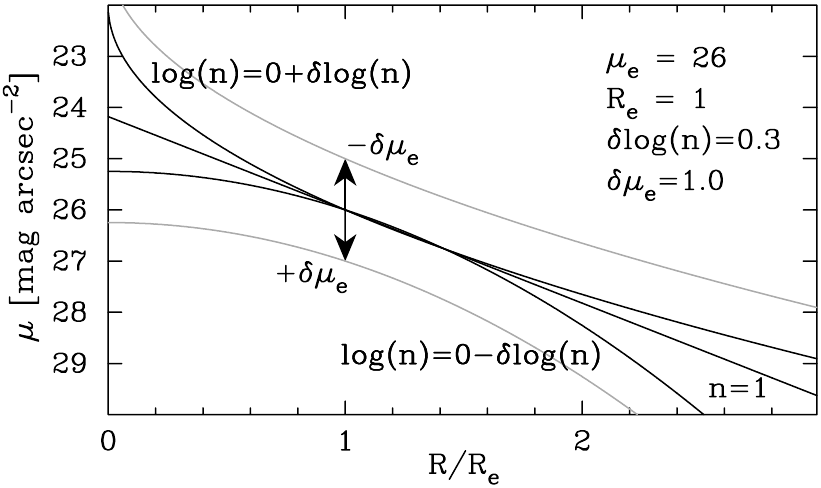}
\caption{An example of how variations/offsets in $\mu_{\rm e}$ and $n$ for a
  surface brightness profile will result in the offset to $\mu_{\rm 0}$.  For
  a given $\mathfrak{M}$, with an associated $\mu_{\rm 0}$ and $n$ from the
  $\mathfrak{M}$-$\mu_{\rm 0}$ and $\mathfrak{M}$-$\log(n)$ relations, the
  horizontal offsets $\delta\mu_{\rm 0}$ seen in
  Figures~\ref{Fig1}a--\ref{Fig3}a are attributable to the offsets
  $\delta\mu_{\rm e}$ and $\delta\log(n)$.  }
\label{Fig4}
\end{center}
\end{figure}

\begin{figure}[ht]
\begin{center}
\includegraphics[angle=0, width=1.0\columnwidth]{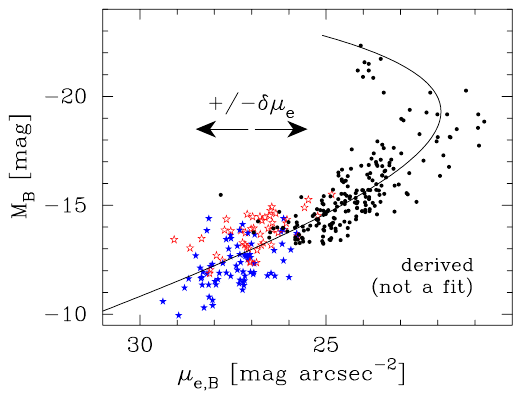}
\caption{Effective surface brightness, $\mu_{\rm e,B}$, at $R_{\rm e}$ versus
  $\mathfrak{M}_B$.
Symbols have the same meaning as in Fig~\ref{Fig2}. 
  The curve is not a fit but instead a derivation from the
  $\mathfrak{M}_B$-$\mu_{\rm 0}$ and $\mathfrak{M}_B$-$\log(n)$ relations
  (Equations~\ref{Eq_M-mu2} and \ref{Eq_Mn}). 
}
\label{Fig5}
\end{center}
\end{figure}

Understanding the intrinsic scatter in the scaling relations is beneficial for
interpreting where UDGs and NUDGes sit relative to the brighter population of
ETGs. Before assessing how low-surface-brightness systems deviate from or
extend these relations, it is helpful to examine the causes and implications
of scatter across the full mass spectrum. In particular, the high-mass end of
ETGs offers key insight into how systematic trends and physical diversity
influence relation slopes and scatter, concepts equally important when
analysing the faint, diffuse regime occupied by UDGs.

Dry major mergers between ETGs containing supermassive black
holes can scour away the core of the newly-formed galaxy, producing the abrupt
turn in the $\mathfrak{M}_B$-$\mu_{\rm 0,B}$ relation at $\mathfrak{M}_B\approx-20.5$~mag 
\citep{2003AJ....125.2936G} 
  rather than at the previously suggested
  $\mathfrak{M}_B\approx-18$~mag \citep[][see their discussion of their Figure~4c based on
  the nucleated and stripped galaxy M32]{1985ApJ...295...73K,
  1997AJ....114.1771F}. 
  \citet{2023MNRAS.520.1975G} suggested that extensive scouring
  from the multiple mergers
\citep[aided by captured satellites:][]{2010ApJ...725.1707G, 2015ApJ...807..136B} 
  that build brightest cluster galaxies, typically massive ETGs, 
leads to such a dramatic
redistribution of stars that the galaxy's reduced central concentration and stellar
build-up at larger galactic radii changes the galaxy profile from having an
obvious core, and thus a core-S\'ersic light profile, to a low-$n$ S\'ersic
profile that may additionally have an exponential-like envelope 
\citep[e.g.][]{2007MNRAS.378.1575S}. This transformation was termed `galforming'. 
While this explains large-scale departures of E galaxies at high masses from
the $\mathfrak{M}_B$-$\mu_{\rm 0,B}$ and $\mathfrak{M}_B$-$\log(n)$ relations,
the scatter about these relations can be 
mined for insight into the formation and evolution of ETGs.
As displayed in Figure~\ref{Fig3}, 
$\delta \mu_{\rm 0,B}$ and $\delta \log(n)$ shall denote the horizontal scatter/offset
between the data and the `expected' value
(from the $\mathfrak{M}_B$-$\mu_{\rm 0,B}$ and $\mathfrak{M}_B$-$\log(n)$
relations) at the same magnitude.

\begin{figure*}[ht]
\begin{center}
\includegraphics[angle=0, width=1.0\textwidth]{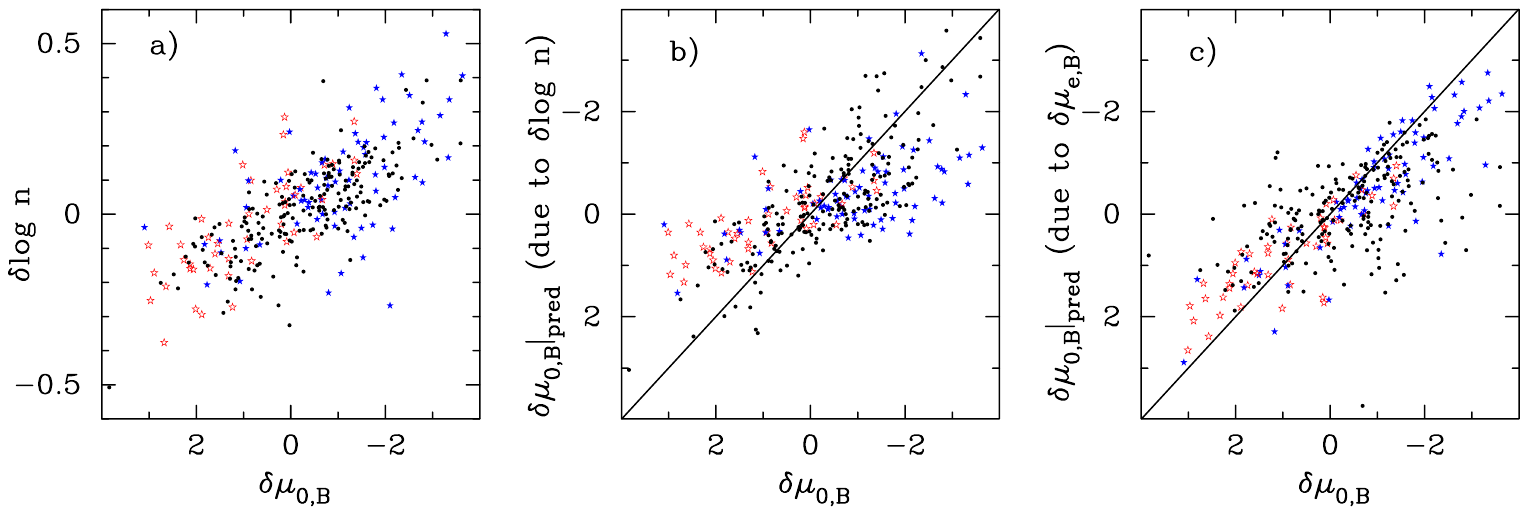}
\caption{Breaking down the scatter in the $\mathfrak{M}_B$-$\mu_{\rm 0,B}$ diagram for ETGs.
Symbols have the same meaning as in Fig~\ref{Fig2}.   
  At fixed $\mathfrak{M}_B$, the horizontal scatter $\delta\mu_{\rm 0,B}$ and
  $\delta\log(n)$ from Figures~\ref{Fig2}a) and ~\ref{Fig2}b) are shown
  against each other in panel a).  The predictable contribution to $\delta\mu_{\rm 0,B}$
  arises from offsets in $\log(n)$ and $\mu_{\rm e,B}$ from their expected values
  (based on the scaling relations).  These are shown in panels b) and c), 
  respectively. The different behaviour of the UDGs from the (dwarf and
  ordinary) ETGs is explained in Section~\ref{Sec_mind} and simply reflects their different 
  S\'ersic indices.
}
\label{Fig6}
\end{center}
\end{figure*}

One can deduce a lot from 
coupling the $\mathfrak{M}_B$-$\mu_{\rm 0,B}$ and $\mathfrak{M}_B$-$\log(n)$ relations with the properties of the
3-parameter S\'ersic model used to describe ETG light profiles. 
The previous section showed how 
$\mathfrak{M}_B$, $\mu_{\rm 0,B}$, and $n$ dictate the value of $R_{\rm e}$ and $\langle\mu\rangle_{\rm e,B}$
(and $\mu_{\rm e,B}$). 
Conversely, $R_{\rm e}$, $\mu_{\rm 0,B}$ (or $\langle\mu\rangle_{\rm e,B}$
or $\mu_{\rm e,B}$), and $n$ yield the value of $\mathfrak{M}_B$ for a given
S\'ersic light profile. 
Here, it is revealed how the scatter about the $\mathfrak{M}_B$-$R_{\rm e}$ relation
can be traced back to the scatter about the $\mathfrak{M}_B$-$\mu_{\rm 0,B}$ relation, which is
directly related to the galaxies' variation in both $\mu_{\rm e,B}$ and $n$ at a given
magnitude (see Figure~\ref{Fig4}).

Equation~\ref{Eq_step} reveals that, in a sense, a galaxy's values of $\mathfrak{M}$ and
$R_{\rm e}$ depend on its value of $\langle\mu\rangle_{\rm e}$, and Equation~\ref{Eq_mumu}
shows how $\langle\mu\rangle_{\rm e}$ depends on $\mu_{\rm e}$ and $n$.  Therefore, for a
given $\mathfrak{M}$, the value of $R_{\rm e}$ depends on $\mu_{\rm e}$ and $n$. 
Consequently, at a given $\mathfrak{M}$, the scatter about the $\mathfrak{M}$-$R_{\rm e}$ curve
relates to the scatter in $\mu_{\rm e}$ and $n$ about their median expected
value, as traced by the $\mathfrak{M}$-$\mu_{\rm e}$ and $\mathfrak{M}$-$\log(n)$
relations.
The latter horizontal scatter is seen in the $\mathfrak{M}$-$\log(n)$
diagram (Figure~\ref{Fig3}b), while the former horizontal scatter is displayed
in Figure~\ref{Fig5}.

For each galaxy, with its observed values of $\mathfrak{M}_B$ and $n$, one additionally has the expected
value of $n$ from the $\mathfrak{M}_B$-$\log(n)$ relation. One can then
calculate $\mu_{\rm 0,B} -
\mu_{\rm e,B}$ for each of these two values of $n$ (Equation~\ref{Eq_meow}) in
order to explore how the difference/offset in the 
S\'ersic index, $\delta \log(n)$, propagates into a horizontal offset of central surface
brightness about the $\mathfrak{M}_B$-$\mu_{\rm 0,B}$ relation. 
The remaining scatter about the $\mathfrak{M}_B$-$\mu_{\rm 0,B}$ relation is due to the
difference between the expected value of $\mu_{\rm e,B}$ (based on the expected
values of $\mu_{\rm 0,B}$ and $n$ for any $\mathfrak{M}_B$) and the observed value of
$\mu_{\rm e,B}$.
These offsets are denoted $\delta \mu_{\rm e}$, with an example
illustrated in Figure~\ref{Fig4} using $n=1$ and $\mu_{\rm e} = 26$ mag
arcsec$^{-2}$ as the `expected' values.  From Figure~\ref{Fig4}, one can envisage how changes to $n$ and
$\mu_{\rm e}$ alter $\mu_{\rm 0}$.  

Figure~\ref{Fig6}a reveals that, for a given $\mathfrak{M}_B$, UDGs with a brighter
$\mu_{\rm 0,B}$ than `expected' (from the $\mathfrak{M}_B$-$\mu_{\rm 0,B}$ relation) also have a higher
value of $n$ (than expected from the $\mathfrak{M}_B$-$\log(n)$ relation). In contrast, those with a
fainter-than-expected value of $\mu_{\rm 0,B}$ have a lower-than-expected value of $n$
for their absolute magnitude.  This change in the S\'ersic index
accounts for some of the horizontal scatter in the $\mathfrak{M}_B$-$\mu_{\rm
  0,B}$ diagram, as shown in Figure~\ref{Fig6}b. 
The offset in $\mu_{\rm 0,B}$ due to the differences between the observed and
expected values of $n$ is given by
$(2.5/\ln(10))[b_{\rm obs} - b_{\rm expected}]$, with $b\approx 1.9992n-0.3271$,
for $0.5<n<10$ \citep{1989woga.conf..208C}. 
Figure~\ref{Fig6}b also reveals that,  
for UDGs with their small S\'ersic indices, the offset
in the observed value of $n$ from the expected value accounts for about
one-third
of the UDGs' offset between the observed value of $\mu_{\rm 0,B}$ and the expected value at that
magnitude.  For brighter ETGs, with their larger values of $n$,
the difference between the observed and expected value of $n$ accounts for
most of their scatter about the $\mathfrak{M}_B$-$\mu_{\rm 0,B}$ relation. 
This is as expected given how the difference between
$\mu_{\rm e}$ and $\mu_{\rm 0}$ in the $R^{1/n}$ model increases roughly
linearly with the S\'ersic index
while, and this is the key, 
the logarithmic scatter $\delta \log(n)$ about the $\mathfrak{M}_B$-$\log(n)$ relation
does not sufficiently decline as $n$ increases. That is, the horizontal scatter
$\delta n$ about the $\mathfrak{M}_B$-$\log(n)$ relation increases with increasing $n$. 

The difference
between the observed value of $\mu_{\rm e,B}$ and the expected value at each
magnitude (see the curve in Figure~\ref{Fig5}) accounts for the remaining offset of the observed value of $\mu_{\rm 0,B}$
from the expected value of $\mu_{\rm 0,B}$ at that magnitude, i.e.\ the scatter. This component of the
difference in $\mu_{\rm 0,B}$ is shown in Figure~\ref{Fig6}c, where it can be seen
that differences in $\mu_{\rm e,B}$ dominate the difference in $\mu_{\rm 0,B}$ for the
UDGs.  For the (brighter) ETGs, the contribution to the scatter about the
$\mathfrak{M}_B$-$\mu_{\rm 0,B}$ relation stemming from the different
$\mu_{\rm e,B}$ values at a given $\mathfrak{M}_B$ is
seen to be sub-dominant to the scatter arising from the difference between
their observed and expected S\'ersic index.

In essence, the situation for the UDGs can be summarised as follows.
UDGs with $\mu_{\rm 0,B}$ values fainter than expected for their magnitude
(based on the $\mathfrak{M}_B$-$\mu_{\rm 0,B}$ relation) and with S\'ersic
indices smaller than expected (based on the $\mathfrak{M}_B$-$\log(n)$
relation) will have effective half-light radii that are larger than expected,
where `expected' means the value coming from the $\mathfrak{M}_B$-$\mu_{\rm
  0,B}$ and $\mathfrak{M}_B$-$\log(n)$ relations. 
Equally, UDGs with brighter than expected $\mu_{\rm 0}$ values and larger than
expected $n$ values will have smaller than expected $R_{\rm e}$ values.  These
shifts (in $R_{\rm e}$) from expectation come on top of the curved
$\mathfrak{M}_B$-$R_{\rm e}$ relation for ETGs, of which UDGs are a part.

\subsubsection{Ellipticity}\label{Sec_ell}

\begin{figure}[ht]
\begin{center}
\includegraphics[angle=0, width=1.0\columnwidth]{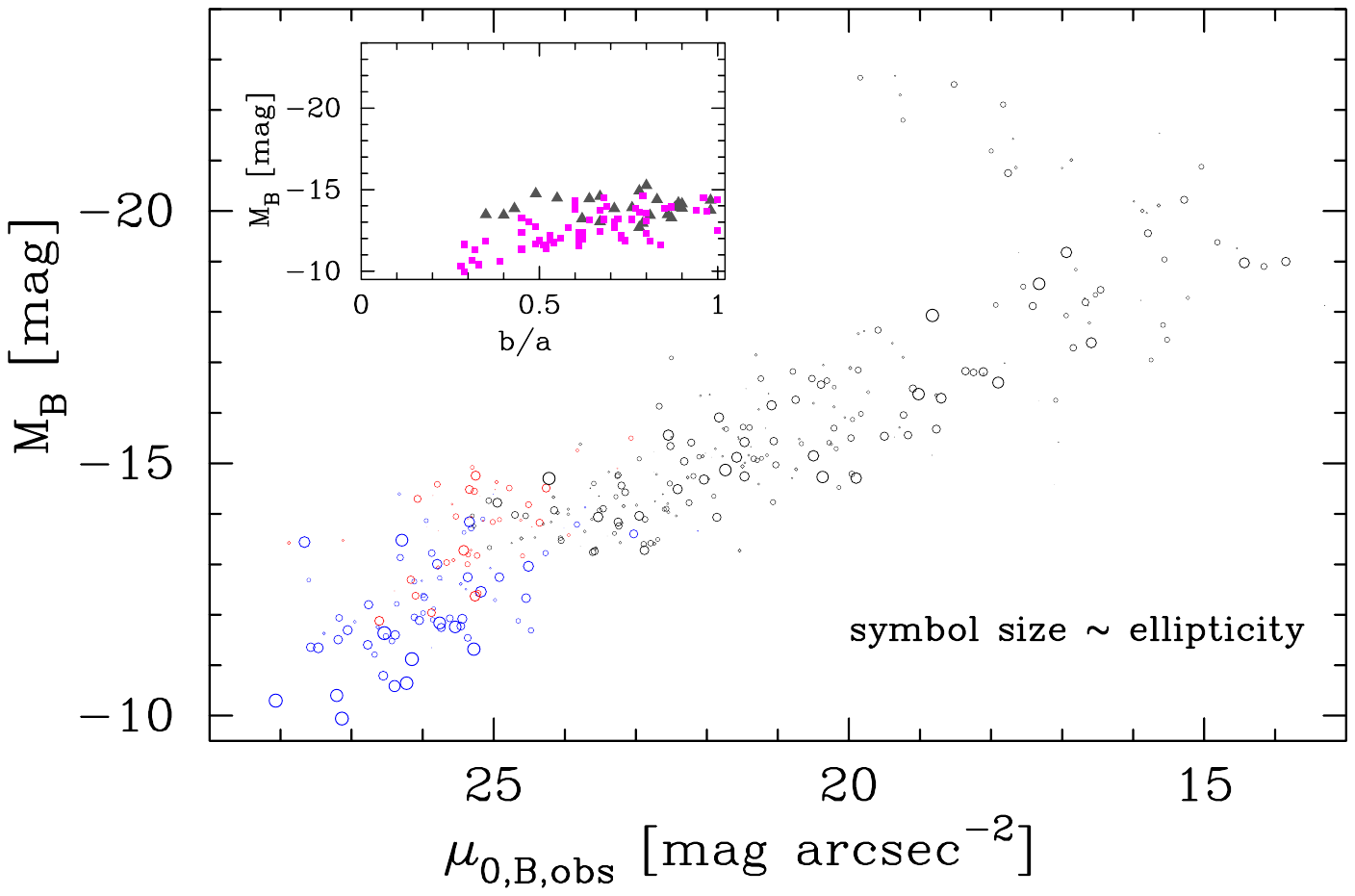}
\caption{Variant of Figure~\ref{Fig2}a, in which the symbol size is now
  proportional to the ellipticity ($=1-b/a$) such that galaxies that appear
  round have a small symbol size.  In the inset panel, 
  the grey triangles represent 28 UDGs predominantly in clusters, while the pink
squares represent 59 UDGs in low-to-moderate density environments.
  UDGs with high $b/a$ ratios are prevalent in both environments and are thus not a product
  of the environment.   At faint absolute magnitudes, 
  the samples' $b/a$ ratios are small, possibly a result of
  sample selection (see the discussion in Section~\ref{Sec_ell}).
}
\label{Fig7}
\end{center}
\end{figure}

A factor that could contribute to some of the horizontal scatter about the
$\mathfrak{M}_B$-$\mu_{\rm 0,B}$ relation is the line-of-sight through the
ETGs, with longer sight lines potentially leading to brighter surface
brightnesses.  It is important to understand that this can arise in two
distinct ways.  If the (dust-free) ETGs are oblate or disc-like, then those
ETGs viewed with a more edge-on orientation will appear to have higher
ellipticities and brighter surface brightnesses.  As such, one would expect to
find a predominance of low $b/a$ galaxies occupying the right-hand side (high
surface brightness side) of the $\mathfrak{M}_B$-$\mu_{\rm 0,B}$ distribution.
If, on the other hand, ETGs are more prolate, with one long axis and two
shorter axes, then the sightline depth through these galaxies will be greatest
when the long axis aligns with our line-of-sight and the $b/a$ axis ratio is
highest. Conversely, their surface brightness will be faintest when they
appear most elongated, i.e.\ higher ellipticity.
Figure~\ref{Fig7} displays the ellipticity of the ETGs.  There is no obvious
trend for high-ellipticity galaxies to occupy either the left or right side of
the $\mathfrak{M}_B$-$\mu_{\rm 0,B}$ distribution.

Cluster dETGs might be vertically-heated discs, which need not have ever
hosted a spiral pattern.  Should clusters be harassing disc galaxies to make
them smaller and more ellipsoidal \citep{1996Natur.379..613M}, then the effect
should be greater on the lower mass systems and, thus, there should be a
difference between the mean ellipticity of field and cluster UDGs of the same
absolute magnitude.  \citet{2020ApJ...899...78R} reported that UDGs near the
centres of clusters are rounder than those near the outskirts, but those near
the outskirts were also fainter, thus convoluting any attribution to an
environmental effect because this $\mathfrak{M}$-$\epsilon$ trend is also seen
with UDGs not in clusters (Figure~\ref{Fig7}).

\citet{2025MNRAS.536.2536B} report that the sample of 59 UDGs from
\citet{2024MNRAS.529.3210B} reside in low-to-moderate density environments,
whereas galaxies in clusters heavily dominate the remodelled sample from
\citet{2022MNRAS.517.2231B}.  The inset panel in Figure~\ref{Fig7} shows the
range of observed $b/a$ axis ratios for these two UDG samples. It is
informative.  There is no clear environmental dependence for this sample.  High $b/a$ values
are observed in the field and cluster ETGs, ruling out an environmental driver
for this sample's shapes. What is also apparent is that at faint magnitudes, the ratios
are low, that is, at faint magnitudes the ellipticity is high. If these are
oblate disc-like galaxies at faint magnitudes, then some of this population
should have a relatively face-on orientation and high $b/a$ ratio. Given the
lack of such detections, it seems plausible that sample selection effects are
at play such that at these faint magnitudes, only systems whose surface
brightness has been enhanced by an increased line-of-sight depth have made it
into the sample.
Finally, if ordered rotation had produced oblate discs in the UDGs, impeding
their initial gravitational collapse and resulting in larger sizes, then the
larger UDGs would be associated with a more elliptical, rather than round,
population, which is not observed.

In conclusion, the apparent ellipticity may contribute to some scatter, but it
does not produce strong trends in the present $\mathfrak{M}_B$-$\mu_{\rm 0,B}$
diagram.  It is, however, noted that the brighter S0 galaxies with rather
edge-on discs and thus higher ellipticity may have been filtered from the
sample because, in the past, they were more readily identified as disc
galaxies rather than elliptical galaxies.
Dwarf mass ($M_{\rm \star,gal}/M_\odot \approx $(2-5)$\times10^9$,
$\mathfrak{M}_B \approx $$-$17.5$\pm0.5$ mag) ETGs in clusters have the same
kinematic properties as those in isolated field environments
\citep{2017MNRAS.468.2850J}.  This further undermines the notion that
cluster-related processes have had a strong hand in shaping dETGs, implying that they are
born as they are rather than transformed. 
This 
was the conclusion reached by \citet{1986AJ.....91...70V} in the case of IC~3475.

\subsection{Isophotal radii}

\begin{figure}[ht]
\begin{center}
\includegraphics[angle=0, width=1.0\columnwidth]{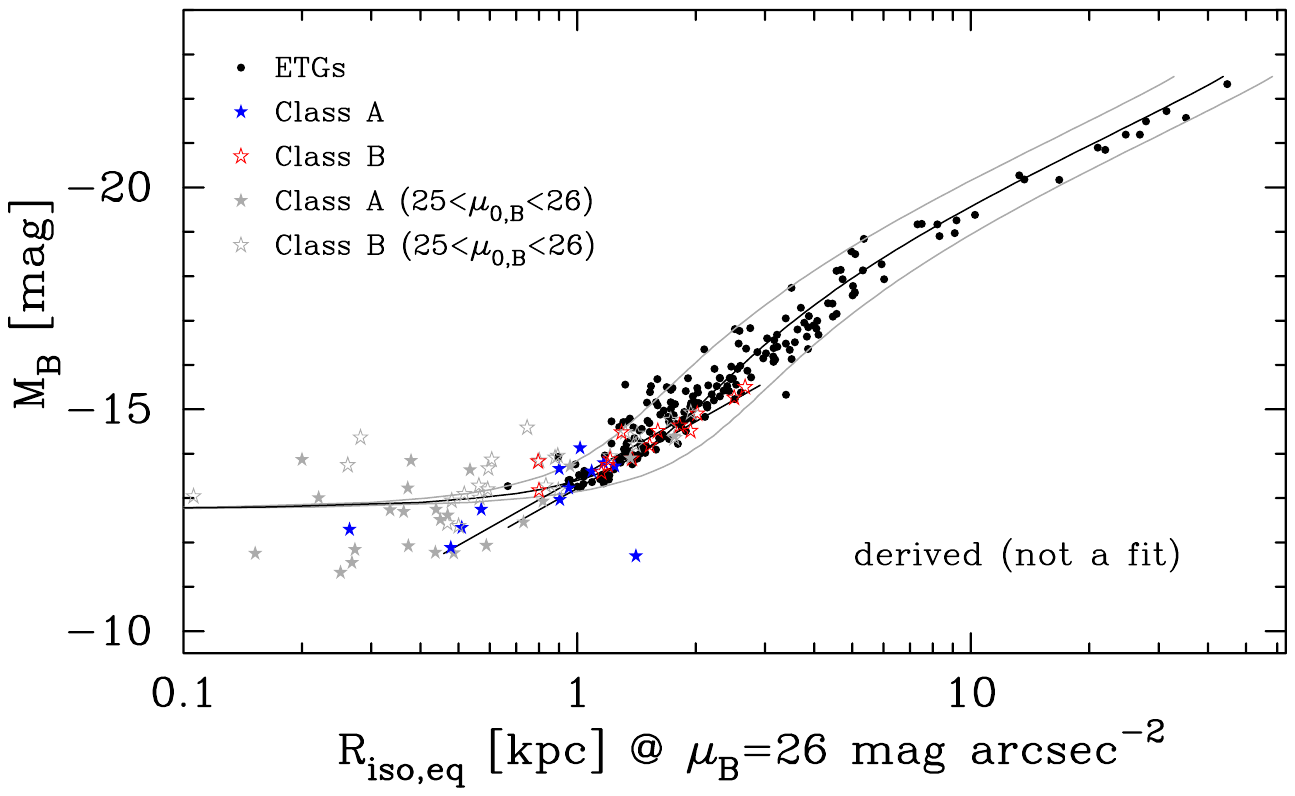}
\caption{Absolute magnitude versus the `equivalent axis' isophotal radius at
  $\mu_{\rm B}=26$~mag arcsec$^{-2}$ obtained from each galaxy's S\'ersic profile.  The central
  black curve is derived using Equations~\ref{Eq_Mn} and \ref{Eq_M-mu2}, while
  the outer grey curves denote a $\pm0.125$~dex offset in isophotal radius at
  fixed magnitude. 
  The two diagonal lines pertain to a fixed $B$-band central surface brightness of
  24 and 25 mag arcsec$^{-2}$ and reveal how $R_{\rm iso}$ changes if $\mathfrak{M}_B$
  changes by up to $\pm$1.6~mag while the associated S\'ersic index is held
  fixed (at $n=0.986$ and 0.861).
  The faint/grey symbols in the lower-left have $\mu_{\rm 0,B} > 25$
  mag arcsec$^{-2}$.  The outlier in the lower-middle is MATLAS-1408. 
} 
\label{Fig8}
\end{center}
\end{figure}

Following \citet{1991AnA...252...27B},
who used ETGs and UDGs from \citet{1985AJ.....90.1681B}, 
Figure~\ref{Fig8} shows the geometric-mean ($R_{\rm eq}$)
isophotal radii where $\mu_{\rm B}=26$ mag arcsec$^{-2}$ versus the $B$-band
absolute magnitude.\footnote{For reference, the Third Reference Catalogue 
\citep[RC3:][]{1991rc3..book.....D} used $\mu_{\rm B}=25$ mag arcsec$^{-2}$, 
while \citet{1969ApL.....3...19H} used $\mu_{\rm V}=26$ mag arcsec$^{-2}$.}
As \citet[][their Figure~2, with
  $-20 < \mathfrak{M}_B < -12$~mag]{1991AnA...252...27B}
demonstrates, see also
\citet[][their Figure~7, with $\mathfrak{M}_B < -14.5$~mag]{1994ApJS...93..397C},
the IC~3475 type galaxies have smaller isophotal radii than brighter ETGs. 
As shown since then, for example, 
\citet[][their Figure~3]{2008MNRAS.389.1924F} and 
\citet[][their Figure~1]{2014PASA...31...11S}, and using (stellar
mass)-density profiles 
\citep{2020A&A...633L...3C, 2020MNRAS.493...87T}, 
the $\mathfrak{M}_B$-(isophotal radius, $R_{\rm iso}$) 
relation \citep{Heidmann67, 1969ApL.....3...19H} 
also has less scatter than the curved $\mathfrak{M}_B$-$R_{\rm \rm e}$
relation \citep[e.g.][]{2006AJ....132.2711G}. 
For the first time, this reduced scatter is explained here.
The slight curvature in the $\mathfrak{M}_B$-$R_{\rm iso}$ relations for ETGs
has been known for decades, and it remains when using a range of different
deprojected (internal) luminosity densities \citep[][his
  Figure~11]{2019PASA...36...35G}.  This is somewhat different to the
magnitude/mass-isodensity relations involving the stellar
edges of thin spiral galaxy discs \citep[e.g.][and references
  therein]{2024ApJ...974..247C}, and conflating the two may hide key 
insight. 

First, however, it is shown how 
the UDGs with $\mu_{\rm 0,B} \gtrsim 25$ mag arcsec$^{-2}$
tend to be impacted by the above isophotal level ($\mu_{\rm B}=26$ mag
arcsec$^{-2}$), which is rather close to their 
central surface brightness value. This proximity causes some UDGs
to scatter to small radii, as shown by the
faint/grey symbols in Figure~\ref{Fig8}.
Given that $\mu_{\rm e}-\mu_{\rm 0}=1.822$ for an $n=1$ profile, one can
immediately appreciate how such isophotal radii will be small for galaxies
with $\mu_{\rm 0} \gtrsim 25$ mag arcsec$^{-2}$ because these isophotal
radii barely sample the galaxy. 
Furthermore, those UDGs with $\mu_{\rm 0,B} > 26$ mag
arcsec$^{-2}$ will have a meaningless isophotal radius of zero. 
Obviously, isophotal levels that are too
bright, or stellar mass density levels that are too high, will be problematic.
For example, the UDG MATLAS-1302 has $\mu_{\rm 0,B}=28.11$ mag arcsec$^{-2}$,
assuming $B_{\rm Vega}=g_{\rm AB}+0.47$.
For reference, adopting a $B$-band stellar mass-to-light ratio of 2.5 gives a 
central stellar mass surface density of 0.93 $M_\odot$ pc$^{-2}$. 
Therefore, a fainter than traditionally-used isophotal level will additionally
be included below. 

Recently, \citet{2024MNRAS.535..299G} has revealed how galaxy mergers have led
to the anti-truncation of discs in dust-rich merger-built S0 galaxies and,
more generally, to the S\'ersicification of massive ETGs. The extended nature
of high-$n$ galaxy light profiles at large radii is explained by the
superposition of acquired progenitor galaxy `discs' and the accumulation of
accreted stellar material and stars flung to large radii.  Such mergers, and
tidally heated discs or UDGs, can push a galaxy's outer stellar density below
a gas disc's threshhold density for star formation, thereby requiring rather
faint isophotal levels or projected stellar densities to measure their sizes.

\begin{figure}[ht]
\begin{center}
\includegraphics[angle=0, width=1.0\columnwidth]{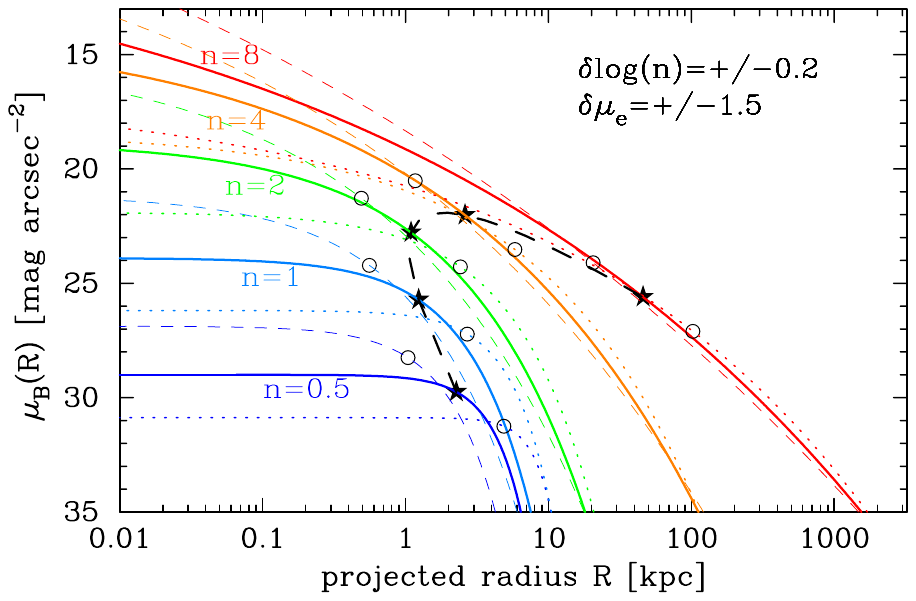}
\caption{Five representative $B$-band light profiles of ETGs with different
  S\'ersic indices are shown (thick solid curves).  For each value of $n$, an
  associated absolute magnitude and central surface brightness are assigned
  from Equations~\ref{Eq_Mn} and \ref{Eq_M-mu2}, from which the effective
  surface brightness and radius can then be calculated
  (Equation~\ref{Eq_Mag_R}).  For each profile, the `effective' parameters
  $\mu_{\rm e,B}$ and $R_{\rm e}$ are shown by the stars, and the dashed curve
  connecting them traces the $\mu_{\rm e,B}$-$R_{\rm e}$ relation for ETGs.
  While keeping the absolute magnitude fixed, offsets of $\delta \mu_{\rm e} =
  \pm1.5$ mag arcsec$^{-2}$ and $\delta\log(n)=\pm0.2$~dex are applied to each
  profile, thereby generating offsets in $\mu_{\rm 0,B}$.  This produces the
  dotted and dashed curves, whose $\mu_{\rm e}$ and $R_{\rm e}$ values are
  shown by the open circles.  These offset profiles yield slight differences
  in the isophotal radii relative to the larger changes seen in $R_{\rm e}$,
  provided the chosen isophotal surface brightness is not too close to the
  light profile's central surface brightness. }
\label{Fig9}
\end{center}
\end{figure}

Before proceeding, 
the $\mathfrak{M}_B$-$\mu_{\rm 0,B}$ and $\mathfrak{M}_B$-$\log(n)$ relations
(Equations~\ref{Eq_M-mu2} and \ref{Eq_Mn}) are used to create five 
representative light profiles of the ETGs and UDGs with values of $n=0.5$, 1,
2, 4, and 8.  The thick curves in Figure~\ref{Fig9} show these profiles, and
their $\mu_{\rm e,B}$ and $R_{\rm e}$ values are marked with a star.  Joining
these stars traces out the curved $\mu_{\rm e,B}$-$R_{\rm e}$ relation for
ETGs and UDGs.  Next, accompanying each of these five light profiles are two
offset profiles with the same $\mathfrak{M}_B$.  One has had
$\delta\log(n)=0.2$~dex and $\delta\mu_{\rm e}=-1.5$ mag arcsec$^{-2}$
applied, while the other has had $\delta\log(n)=-0.2$~dex and $\delta\mu_{\rm
  e}=+1.5$ mag arcsec$^{-2}$ applied.  Their offset values of $\mu_{\rm e,B}$
and $R_{\rm e}$ are calculated and marked with open circles. 
For each of the five cases,
one can immediately see that the change to the isophotal radii from these
perturbations is smaller than the change to $R_{\rm e}$.  That is, the scatter
at fixed $\mathfrak{M}_B$ about the $\mathfrak{M}_B$-$\mu_{\rm 0,B}$ (and
$\mathfrak{M}_B$-$\log(n)$) relations will create a larger scatter about the
$\mathfrak{M}_B$-$R_{\rm e}$ relation (Figure~\ref{Fig2}c) than it will about
an $\mathfrak{M}_B$-$R_{\rm iso}$ relation (unless the isophotal level is
close to the central surface brightness of the light profile, as witnessed in
Figure~\ref{Fig8}).

\begin{figure}[ht]
\begin{center}
\includegraphics[angle=0, width=1.0\columnwidth]{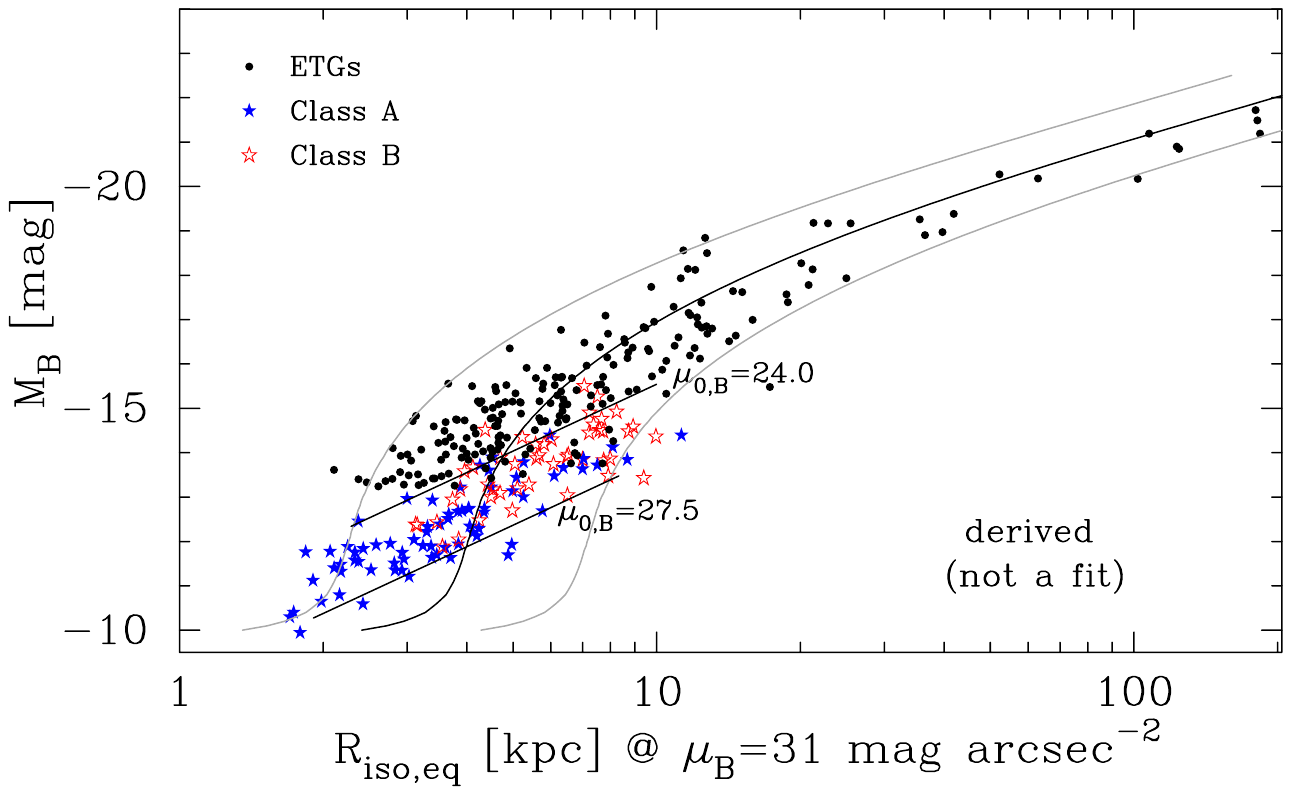}
\caption{Absolute magnitude versus the `equivalent axis' isophotal radius at
  $\mu_{\rm B}=31$ mag obtained from each galaxy's S\'ersic profile.  The central
  black curve is derived using Equations~\ref{Eq_Mn} and \ref{Eq_M-mu2}, while
  the outer grey curves denote a $\pm0.25$~dex offset in isophotal radius at
  fixed $\mathfrak{M}_B$. The two diagonal lines pertain to a fixed $B$-band central surface brightness of
  24 and 27.5 mag arcsec$^{-2}$ and reveal how $R_{\rm iso}$ changes if $\mathfrak{M}_B$
  changes by up to $\pm$1.6~mag while the associated S\'ersic index is held fixed (at
  $n=0.986$ and 0.613, respectively).} 
\label{Fig10}
\end{center}
\end{figure}

To circumvent the above problem for UDGs that an isophotal limit of
$\mu_{\rm B}=26$ mag (and $\mu_{\rm B}=27$--28 mag) causes, 
the $\mathfrak{M}_B$-$\mu_{\rm 0,B}$ and $\mathfrak{M}_B$-$\log(n)$ relations
have been used to 
derive the `expected' isophotal radius where $\mu_{\rm B}=31$ mag arcsec$^{-2}$ for different values
of $\mathfrak{M}_B$. 
This is shown in Figure~\ref{Fig10} by the thick curve. For each galaxy from 
the data samples, which are overplotted in this figure, 
their S\'ersic parameters have been used to compute the radius where $\mu_{\rm B}=31$ mag
arcsec$^{-2}$.  
That is, the calculations assume that the galaxies' S\'ersic profiles can be extrapolated
this far.
This is a very faint surface brightness level, and where 
triaxial and disc galaxies actually end is a separate topic 
\citep{2014ApJ...784..142S, 2022A&A...667A..87C}. Additional caveats beyond the scope of the current
analysis include departures in the light profiles arising from
truncated and anti-truncated discs in some ETGs \citep{2015A&A...580A..33E, 2024MNRAS.535..299G},
tidal debris streams \citep{2005AJ....130.2647V, 2010ApJ...715..972J,
  2024A&A...691A.196M}, and 
cluster light creating envelopes around brightest cluster galaxies
\citep{2007MNRAS.378.1575S, 2008A&A...483..727P}. In practice, use of a
brighter isophotal level may prove desirable. 
The Large Synoptic Survey Telescope ({\it 
LSST})\footnote{\url{https://www.lsst.org/}} \citep{2019ApJ...873..111I}
will probe most galaxies to new
depths, and the European Space Agency's ARRAKIHS (Analysis of Resolved
Remnants of Accreted galaxies as a Key Instrument for Halo Surveys) 
mission\footnote{\url{https://www.arrakihs-mission.eu/}} will reach a surface
brightness of 31 AB mag arcsec$^{-2}$ \citep{2022eas..conf.1507G}. Additional
upcoming facilities capable of probing the very outskirts of galaxies include 
the Nancy Grace Roman Space Telescope \citep[NGRST:][]{2013arXiv1305.5422S,
  2019arXiv190205569A} 
and the Extremely Large Telescope \citep{2023ConPh..64...47P}.

For galaxies with thin discs, such as spiral galaxies, the outer surface
brightness is closely tied to the local gas volume density (e.g. from
H{\footnotesize I} or H{\footnotesize II}), and hence to the star formation
threshold. This is because in such systems, the line of sight through the disc
is short and well-defined, so the observed surface brightness is a good proxy
for the local stellar and gas volume densities.  In contrast, for early-type
galaxies (ETGs) that have triaxial or thick oblate/prolate stellar
distributions, the relationship is more complex. The surface brightness in
these galaxies represents the integrated light along extended lines of sight
through the galaxy’s volume, making it a poor tracer of the local volume
density at any given point. To estimate the internal density distribution in
these systems, one must perform deprojections 
\citep[e.g.][]{1991A&A...249...99C} --- which differ from simple inclination
corrections --- in order to recover the three-dimensional structure.  This has
direct implications for UDGs with triaxial shapes: their low surface
brightness is often a result of geometric projection, and is not
indicative of a local stellar density. Therefore, the surface
brightness in such systems may not relate to the conditions needed for star
formation.

Once again, the scatter in the data about the $\mathfrak{M}_B$-$R_{\rm iso}$ curve is
small compared to that of the $\mathfrak{M}_B$-$R_{\rm e}$ curve.  Such
behaviour has been known for decades. 
As with Figure~\ref{Fig2}c, two diagonal lines of constant central surface brightness
have been added to Figures~\ref{Fig8} and \ref{Fig10} to 
illustrate how the UDG/NUDG sample selection again results in a
misleading luminosity-size relation for this sample, 
at odds with that (the thick curved line)
from its parent sample.  That is, the sample selection spawns a relation that deviates from the
trend defined by the ETG population at large.  This same sample selection bias 
was seen to arise in the luminosity-(effective radius) diagram
(Figure~\ref{Fig2}c). 
Clearly, considerable care is needed when interpreting
$\mathfrak{M}_B$-$R_{\rm iso}$ diagrams for UDGs: galaxies defined by an arbitrary
slicing of the broader ETG population.

\section{Discussion}\label{Sec_Disc}

Recent studies have presented a range of definitions for ultra-diffuse
galaxies (UDGs), often anchored to structural criteria such as a minimum
effective radius (e.g.\ $R_{\rm e,maj} \gtrsim 1.5$--1.7~kpc) and a threshold central 
surface brightness (e.g.\ $\mu_{\rm 0,g} \gtrsim 24$ mag arcsec$^{-2}$), as originally outlined by
\citet{2015ApJ...798L..45V} and adopted in similar form by others
\citep[e.g.][]{2015ApJ...807L...2K, 2016ApJS..225...11Y}. 
Additional definitions of UDGs include outliers from linear
\citep{2023ApJ...955....1L} or curved \citep{2020ApJ...899...69L} 
size–luminosity or surface brightness–luminosity relations. 
However such definitions can be 
arbitrary \citep[see][]{2022ApJ...926...92V}
and misleading when applied to a population following a curved trend
with intrinsic scatter. 
Rather than identifying a 
physically distinct class, many of these criteria capture a portion of
the continuous distribution of ETGs in structural parameter
space. Therefore, care must be taken when drawing evolutionary conclusions
based on classifications that may reflect observational bias or parameter
slicing, rather than a genuine bifurcation in formation mechanisms or
intrinsic galaxy properties.
In passing, it is noted that not all UDGs should be considered ETGs, as a small
number of UDGs have spiral structure, particularly among the bluer, gas-rich
UDG population in the field and groups \citep{1987AJ.....94...23B, 2017ApJ...842..133L,
  2017MNRAS.468.4039R, 2017ApJ...846...26S, 2019MNRAS.484.4865P, 2021MNRAS.506.5494P}.
It is also noted that the galaxy sample used here does not include compact
elliptical galaxies like NGC~4486B or M32 --- considered to be the denser,
more spherical bulge component of stripped disc galaxies
\citep{2001ApJ...557L..39B, 2002ApJ...568L..13G}\footnote{The (stellar
mass)-$R_{\rm e}$ relation for bulges/spheroids --- with a galaxy's disc, bar
and other components removed --- can be seen in \citet{2023MNRAS.519.4651H}.
Compact elliptical galaxies, which are essentially the bulge component of
disc-stripped S0 galaxies \citep{2001ApJ...557L..39B, 2002ApJ...568L..13G},
follow this distribution rather than the curved distribution of ETGs.} ---,
nor dwarf spheroidal galaxies \citep{1998ARA&A..36..435M,
  2008ApJ...684.1075M}, with their smaller and denser stellar distributions
than UDGs.  Therefore, their connections to or differences from ETGs and UDGs
are not explored here.

\subsection{Division of ETGs}

The notion that there may be two distinct (dwarf and ordinary) types of ETG
was perhaps first raised by \citet{1968BAICz..19..105S} upon inspecting the
departure of the fainter ETGs from the $\mathfrak{M}_B$--$R_{\rm e}$ relation
defined by the brighter ETGs.  This departure is also seen in the
$\mathfrak{M}_B$--$\langle \mu \rangle_{\rm e,B}$ diagram \citep[e.g.][their
  Figure~3]{1985ESOC...20..239S} and the $\mathfrak{M}_B$--$\mu_{\rm e}$
diagram (Figure~\ref{Fig5}).  However, as detailed in
Paper~I, these departures (and departures in all diagrams
involving `effective' half-light parameters) arise from the systematically
varying -- with absolute magnitude --- \citet{1968adga.book.....S} $R^{1/n}$
nature of their stellar distribution.  The suggestion of
a third (UDG) type of ETG due to their deviation to increasing sizes at yet
fainter magnitudes in the $\mathfrak{M}_B$--$R_{\rm e}$ diagram is dispelled
here because bends in diagrams involving effective half-light parameters
should not be used as a diagnostic tool for galaxy formation.
As
Paper~I explained, the magnitude
associated with the ‘bend point’ in diagrams involving ‘effective’ parameters
changes with the enclosed fraction of light used to define the ‘effective’
parameters.  This voids any physical significance of the bend point in terms
of galaxy formation scenarios.

There are, however, subtypes among the ETGs, which has undoubtedly added to
the confusion when interpreting scaling relations involving effective
half-light parameters.  These subtypes are presented in
\citet{2023MNRAS.522.3588G}
and encompass primordial/primaeval lenticular 
(S0) galaxies --- including the UDGs at the faint end --- that need not have
once been spiral galaxies, plus major-merger-built
S0 galaxies, which are dust-rich if a spiral galaxy was involved in the
collision, ellicular (ES) galaxies with fully embedded discs, and discless
elliptical (E) galaxies.  Therefore, the dwarf-mass ETGs are not simply small
versions of giant S0 or E galaxies in that they represent an earlier stage of
evolution.  This aspect has likely muddied the waters and given false support
to interpretations of a division at $\mathfrak{M}_B \approx -18$ mag based on
scaling diagrams involving $\mu_{\rm e}$, $\langle \mu \rangle_{\rm e}$, and
$R_{\rm e}$.

The yet further division of UDGs and NUDGes into two classes by
\citet{2025MNRAS.536.2536B} can be seen in Figure~\ref{Fig2}.  It is
immediately apparent that Class~B (denoted by the open stars) is, on 
average, brighter than Class~A (denoted by the filled stars). Therefore, the
more massive Class~B will have, on average, redder colours, given the
colour-magnitude relation for dETGs \citep[e.g.][]{1972MmRAS..77....1D,
  2008MNRAS.389.1924F, 2024MNRAS.531..230G}.  The redder
colours will equate to a higher (stellar mass)-to-light ratio, $M_*/L$, for
Class~B, given the well-known $M_*/L$-colour trends
\citep[e.g.][]{2001ApJ...550..212B}.  Class~B will also have higher velocity
dispersions, $\sigma$, given the luminosity-$\sigma$ relation for dETGs
\citep[e.g.][]{1983ApJ...266...41D, 2014ApJS..215...17T}.
Furthermore,
Figure~\ref{Fig2} reveals why the galaxies in
Class~B have larger sizes and, therefore, higher inferred total masses when
using $\sigma^2R_{\rm e}/G$ as a proxy for dynamical mass.  Consequently,
Class~B will have more globular clusters, given how this
scales with the inferred halo mass \citep{2009MNRAS.392L...1S}, modulo the
increased scatter at low-masses \citep{2018MNRAS.481.5592F}.  To summarise,
a division among UDGs (and NUDGes) indirectly but heavily based on
their distribution about the $\mathfrak{M}_B$-$\mu_{\rm 0,B}$ relation will
explain the offsets between the mean value of many parameters for Class~A and
B. This need not imply a second uniquely different origin story \citep[e.g.\
  `puffy dwarfs': ][]{2015MNRAS.449L..46E, 2018RNAAS...2...43C} for LSB ETGs 
unless one is to attribute the scatter about the $\mathfrak{M}_B$-$\mu_{\rm
  0,B}$ relation to different formation scenarios.

The question of whether studies have identified two classes of UDG on either side of
the  $\mathfrak{M}_B$-$\mu_{\rm 0,B}$ relation or whether 
the effect of intrinsic scatter on the scaling relations has been missed 
appears to have been answered in Section~\ref{Sec_anal}.
A greater level of spin, i.e.\ ordered rotation, does not appear to 
correlate with the larger UDGs.  Given that galaxies with a higher spin would
be supported by rotation, in addition to velocity dispersion, one would expect
them to be relatively flatter systems. However, the larger UDGs, which reside on the fainter
central surface brightness side of the $\mathfrak{M}_B$-$\mu_{\rm 0,B}$
relation, do not have smaller axis ratios (Figure~\ref{Fig7}).  Equally, if there is a population
of UDGs with edge-on discs yielding brightened $\mu_{\rm 0}$ values,
Figure~\ref{Fig2}c (and \ref{Fig3}c) require them to have {\it
  smaller} radii (both $R_{\rm eq}$ and $R_{\rm maj}$) than the other UDGs of
the same magnitude.  This is not observed; that is, they are not spun out to larger radii.
What makes the first option
even more problematic is that the scenario to preferentially place those UDGs assumed to have 
`puffed up' after gas loss from stellar winds or ram pressure stripping
\citep{1979PASJ...31..193S, 1986ApJ...303...39D, 1986ApJ...305..669V, 2017MNRAS.466L...1D} on one 
side of the $\mathfrak{M}_B$-$\mu_{\rm 0,B}$ relation seems 
unlikely to preferentially operate on the higher mass UDGs that are more
capable of retaining their gas.
One might expect that the proposed processes to `puff up' some UDGs 
should preferentially operate on fainter, lower mass systems,
yet the brighter, higher mass UDGs have the larger sizes.  This
undermines the relevance of this proposed additional formation scenario. 
Furthermore, the observed size-luminosity trend is readily explained because 
brighter galaxies require larger  sizes (Equation~\ref{Eq_MR}) 
for a given S\'ersic profile with fixed values of $\mu_{\rm 0,B}$ and $n$. 
That is, no recourse to different formation scenarios is required 
to explain the increased sizes or many of the 
trends observed in UDG properties that can be traced back to their differing mass.

Due to the $\sim$3.2~mag of vertical scatter ($\pm2\sigma$) about the
$\mathfrak{M}_B$-$\mu_{\rm 0,B}$ relation, selecting a sample of LSB ETGs
with $24 < \mu_{\rm 0,B} < 27.5$ mag arcsec$^{-2}$ results in an
$\mathfrak{M}_B$-$R_{\rm e}$ distribution that is roughly orthogonal to the
wholesale $\mathfrak{M}_B$-$R_{\rm e}$ correlation for ETGs (Figures~\ref{Fig2}c
and \ref{Fig3}c). This behaviour is also somewhat true for the
$\mathfrak{M}_B$-$R_{\rm iso}$ relation (Figure~\ref{Fig10}).
As \citet{1998ApJ...505..199S} noted, 
these misleading trends of brighter luminosities and larger sizes for samples with
(a small range in) low central surface
brightness had previously been used to separate LSB dwarf galaxies from blue
compact dwarf galaxies.

\begin{figure}[ht]
\begin{center}
\includegraphics[angle=0, width=1.0\columnwidth]{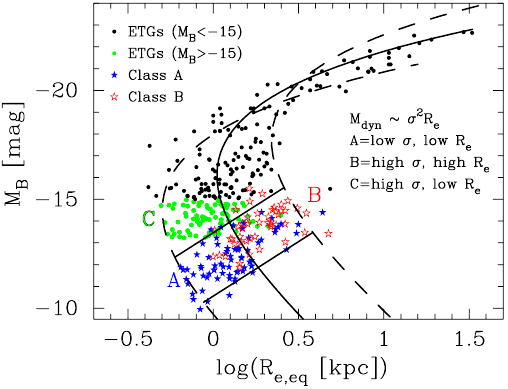}
\caption{Given $\mathfrak{M}_B \propto \sigma^2$ for dETGs \citep[][and
    references therein]{1983ApJ...266...41D, 2013pss6.book...91G}, one can
  appreciate how the scatter about the $\mathfrak{M}_B$-$R_{\rm e}$ relation,
  shown here, 
  results in different dynamical masses ($\sigma^2R_{\rm e}/G$), and in turn, mass-dependent
  trends, that are subject to arbitrary divisions of the dETG (including UDG) sample.  The
  green dots are ETGs not in the UDG sample but with $\mathfrak{M}_B \ge -15$
  mag. As in Figure~\ref{Fig1}c, the curves are not a fit to the data but are derived from
  fits to other empirical relations coupled with the S\'ersic $R^{1/n}$ function.} 
\label{Fig11}
\end{center}
\end{figure}

As seen, faint galaxies
with fainter surface brightnesses than other galaxies of the same faint
magnitude have larger radii than those of the same faint 
magnitude. What causes this is unclear, although it is noted that those with
larger radii also have smaller S\'ersic indices.  Rather than subsequently
being puffed up after formation, 
these large galaxies may initially form this way.
The ETG sequence may well be a primordial and merger-built sequence
representing the long side 
of the Triangal\footnote{The Triangal is a schematic showing the speciation of
galaxies. Unlike galaxy morphology diagrams such as the Tuning Fork
\citep{1928asco.book.....J, 1936rene.book.....H} 
or the Trident \citep{1976ApJ...206..883V}, the Triangal includes evolutionary
pathways, such as major mergers of disc galaxies with strong spirals to form
dust-rich lenticular galaxies.} \citep{2023MNRAS.522.3588G}, encapsulating
primaeval galaxies (UDG/IC~3475, dS0, dE), 
merger-built S0 galaxies, and multiple-merger-built E galaxies, with mass and S\'ersic
index increasing through mergers. 
This notion will be detailed further in a follow-up paper involving kinematical information.
While there are several facets and intricacies to this problem, a broad
overview is presented in Figure~\ref{Fig11}. It shows how the arbitrary and
artificial division of LSB galaxies and regular-size dwarf-mass galaxies will
result in a different average dynamical mass for each subdivision. The
brighter UDGs have higher velocity dispersions due to the
$\mathfrak{M}_B$-$\sigma$ relation, and those with larger $R_{\rm e}$ values
will have bigger dynamical massses when using $\sigma^2R_{\rm e}/G$ as a total
mass indicator, thereby explaining why the dynamical masses of UDG samples
correlates with the richness of their globular cluster system
\citep{2009MNRAS.392L...1S}.  However, they all belong to the ETG population
with a continuous range of properties. Intrinsic scatter about the
$\mathfrak{M}_B$-$\mu_{\rm 0,B}$ and $\mathfrak{M}_B$-$n$ scaling relations
will be addressed in terms of dark matter fractions in a follow-up paper.

Although \citet{2016MNRAS.459L..51A} did not explore if their model for
creating UDGs with larger $R_{\rm e}$ due to a larger angular momentum reproduces
the $\mathfrak{M}_B$-$\log(n)$ relation for ETGs\footnote{While
\citet{2016MNRAS.459L..51A} refer to models having different `concentrations',
they are not referring to a different radial concentration of stars as per the
S\'ersic model \citep{1991A&A...249...99C, 2001MNRAS.326..869T, 2001AJ....122.1707G} but to a
measure of the density contrast of the halo relative to the average background
density of the Universe.}, their conclusion that UDGs need not be shaped by
their environment and that they are the natural extension of ordinary and dwarf ETGs
agrees with this manuscript's alternate analysis of the situation and sample in hand.  The
combination of lower $n$ and fainter $\mu_{\rm 0,B}$ for a given absolute
magnitude will result in a larger $R_{\rm e}$ (Section~\ref{Sec_anal}). This
differs somewhat from the statement by \citet{2016MNRAS.459L..51A} that ``the
UDG population represents the tail of galaxies formed in dwarf-sized haloes
with higher-than-average angular momentum''.  If true, this would necessitate
that, at fixed $\mathfrak{M}_B$, the S\'ersic index is lower (higher) for UDGs
with higher (lower) 
angular momentum. This notion will be pursued in a follow-up paper.

\subsection{Might some UDGs be (precursors to) discs?}\label{Disc_disc}

The stellar discs in many local HSB ETGs
\citep[e.g.][]{1990ESOC...35..231C, 1990ApJ...362...52R}
went overlooked for decades, with
\citet{1998AnAS..133..325G} concluding that the number of (discless) E galaxies
is much lower than previously thought.
This finding supported the view that
low-luminosity\footnote{$\mathfrak{M}_B > -20$ mag, H$_{\rm 0}$=50 km s$^{-1}$ Mpc$^{-1}$.}
ETGs are S0 galaxies \citep{1990ApJ...348...57V}. With 
increased numbers of kinematic maps, it is now known that 
among galaxies that are more massive than $\sim10^9$ M$_\odot$, true pressure-supported
E galaxies
are only found at the high mass end 
\citep[$\gtrsim 10^{11}$ M$_\odot$:][]{2011MNRAS.414..888E,
  2013MNRAS.432.1768K}.  These systems are 
built from the merger and summation of many disc galaxies \citep{2024MNRAS.535..299G},
as required to erase the ordered orbital angular momentum.

Based on the distribution of ellipticities in galaxies deemed, from
visual inspection, to be dE rather than dS0 or Im, it was reported
\citep{Sandage:1980, 1985ESOC...20..239S} that 
they are not discs but rather true elliptical systems. This view has been
pervasive but was partly challenged by \citet{1994ApJ...425...43R}, who
reported a greater flattening than previously recognised. 
\citet{1995AnA...298...63B} subsequently reported  
consistency among the ellipticity distributions of normal (non-nucleated)
dEs, dS0s, late spirals (Sdm - Sms), smooth irregulars (Ims), and clumpy
irregulars (Blue Compact Dwarfs) and a different  distribution for nucleated
dwarfs and a sample of giant ETGs containing ordinary S0 and E galaxies.
For cluster UDGs, \citet{2019MNRAS.485.1036M} report $<b/a>=0.67$ with a
2$\sigma$ scatter (not to be confused with the uncertainty on the mean) of
$\approx\pm0.2$ resembling their expected distribution for thick discs. More
specifically, for UDGs projected within their clusters' $R_{\rm 200}$ radius,
they observe $<b/a>\approx 0.72\pm0.02$ if $R_{\rm e,maj}\lesssim 3.5$ kpc,
and more discy shapes ($<b/a>\approx 0.5$) for the truly giant LSB galaxies not
abundant in the present manuscript's sample.
Based on the $q \equiv b/a$ ratios in \citet[][their Appendix
  B.1 and B.2]{2020ApJ...899...69L} for 44 UDGs in the Coma cluster, they observe a
distribution from 0.3 to 1, consistent with that of dETGs in the Virgo and
Fornax clusters.  They interpret this as evidence of triaxial or spheroidal
shapes rather than preferentially edge-on discs, which would have a flat
distribution ranging from the disc thickness to 1.
While not disagreeing with this, it is remarked that given how the prevalence
of discs --- by which it is meant a substantial, somewhat flattenned, rotating
component --- was missed in many ordinary HSB ETGs for decades, it would not
be surprising if some are still missed in large LSB galaxies.

As with some ordinary ETGs, some LSB dwarf ETGs reveal their disc nature by
displaying bars or spiral patterns \citep[e.g.][]{2000AnA...358..845J,
  2002AnA...391..823B, 2003AJ....126.1787G}.
Initially considered something
of a novelty, the significance of these discoveries went somewhat over-looked
as it was not appreciated how widespread the dS0 rather than dE population
was, although, at that time, it was starting to be discovered that rotation in
dwarf ETGs is common \citep{2002MNRAS.332L..59P, 2003AJ....126.1794G,
  2009ApJ...707L..17T}.
Rotational dominance over velocity dispersion has also been observed in the
stellar kinematics of some UDGs \citet{2025A&A...694A.276B}.
The H{\footnotesize I} gas, when present, is also known to rotate in 
UDGs \citep[e.g.][]{2017ApJ...842..133L}, although at a slower than expected
pace for some \citep{2019ApJ...883L..33M, 2020MNRAS.495.3636M}. 
Another reason why some UDGs are likely to be disc-like rather than spheroidal 
is that the internal (not projected) stellar luminosity density profile derived by
deprojecting the S\'ersic $R^{1/n}$ light profile leads to spheroids with holes in the
middle if $n<0.5$.

Given the declining bulge-to-disc stellar-mass ratio in ETGs with declining
bulge stellar masses down to at least $2\times10^9$ M$_\odot$
\citep[$\mathfrak{M}_B\approx$$-$16 mag:][his Figure~A2]{2023MNRAS.522.3588G},
the continuity in the structural scaling relations shown here implies that
UDGs might also be disc-like galaxies.
Indeed, IC~3475 has a faint bar-like structure \citep{1986AJ.....91...70V,
  1999ApJ...514..119K} and H{\footnotesize I} rotation \citep[$W_{20}=114$ km
  s$^{-1}$:][]{1989A&A...210....1H}, 
revealing its possible disc nature.\footnote{The bar brightens the reported $B$-band
central surface brightness to 23.3 mag arcsec$^{-2}$.}  These latter two studies
struggled to identify what IC~3475 might have evolved from, rejecting the 
proposition that it is a disc-stripped dwarf spiral galaxy. 
More recently, the UDG NGVSUDG-A11 has been found to display a faint spiral
pattern in its stellar distribution \citep{2020ApJ...899...69L}.

The rather common IC~3475 types \citep{1983ApJS...53..375R} may include 
dS0 galaxies with some knotty regions, with some displaying mild irregularities
without apparent dust lanes or clear spiral structure.  Once the irregularities
and knots disperse or settle to the galaxy centre via dynamical friction ---
such sinking is more efficient in the brighter dwarf galaxies
with their higher-$n$ light profile and thus steeper gravitational potentials
within $R_{\rm e}$ \citep{2005MNRAS.362..197T}
--- they will resemble the
so-called dE or nucleated dE,N systems.  These need not be flat discs but
thick (hot) discs that are somewhat rotationally
supported structures, similar to the exponential
components seen in HSB ETG light profiles. 

\begin{figure}[ht]
\begin{center}
\includegraphics[angle=0, width=1.0\columnwidth]{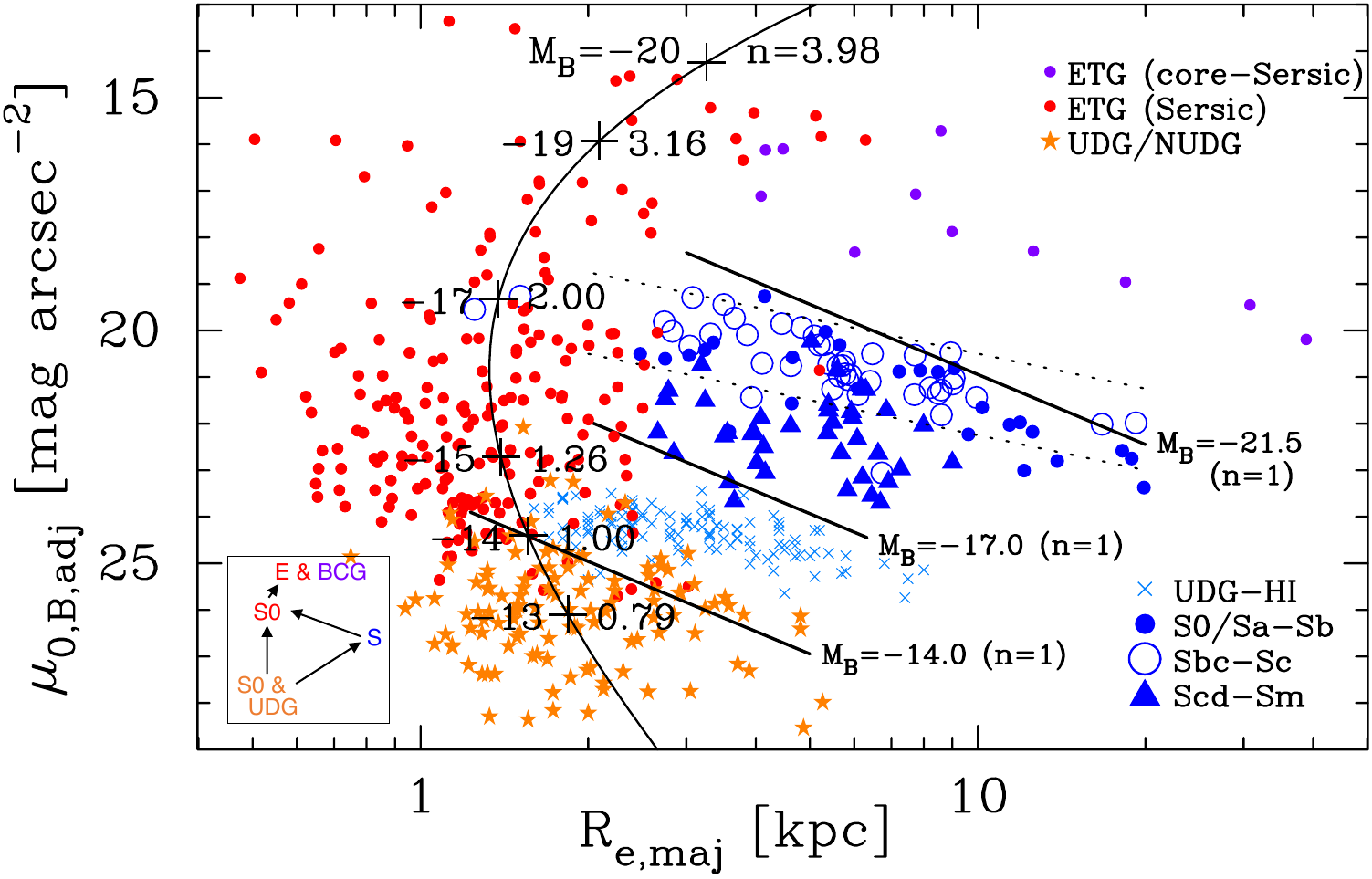}
\caption{ $B$-band (Vega) central surface brightness versus effective half
  light radius for the UDGs, NUDGes and ETGs presented herein
  (Section~\ref{Sec_data}), along with the disc-component of the S0 and spiral
  galaxies compiled by \citet{2001ApJ...556..177G} and the exponential models
  fit to the UDGs and NUDGes detected in neutral hydrogen by
  \citet{2017ApJ...842..133L}.  The curve is not a fit but rather the derived
  expectation following on from Equations~\ref{Eq_Mn} and \ref{Eq_adj}.  Along
  this curve, the location of different absolute magnitudes and the associated
  S\'ersic indices are marked with plus signs.  Lines of constant absolute
  magnitude for exponential ($n=1$) light profiles are also shown, and the
  dotted lines delineate the upper envelope and relation reported by
  \citet{2001MNRAS.326..543G}, replacing the once-popular notion that spiral
  galaxies have a canonical central surface brightness of approximately 21.65
  B-mag arcsec$^{-2}$ \citep{1970ApJ...160..811F}.  The `Triangal'
  evolutionary schematic \citep{2023MNRAS.522.3588G} is illustrated in the
  lower left.  For the relatively gas-poor UDGs and NUDGes depicted there,
  their evolution may occur through two pathways: (i) the acquisition of gas,
  which is collisional and leads to the formation of discs, dissipation, and
  star formation, or (ii) the acquisition of stars (and dark matter),
  resulting in a collisionless, dissipationless merger that results in more massive
  ETG-like system.  The balance between the amount of gas and stars, along with factors
  such as orbital momentum and geometry, likely determines where the needle falls between
  each pathway. }
\label{Fig12}
\end{center}
\end{figure}

\subsection{Further musings}

Explanations, whether through observation, theory, or simulation, for the
$\mathfrak{M}$-$\mu_{\rm 0}$ and $\mathfrak{M}$-$\log(n)$ relations for ETGs
would be welcome.  For the low-$n$ ETGs with faint central surface
brightnesses, including the UDGs presented thus far, it is speculated here that they are
somewhat primordial rather than a transformed population.  They may 
have grown onto the $\mathfrak{M}_B$-$\mu_{\rm 0,B}$ and
$\mathfrak{M}_B$-$\log(n)$ relations, with secondary processes (along with
measurement error and arguably some disc presence and inclination effects)
adding further scatter to these relations, beyond the randomness of accretion
and merger assembly that built the UDGs and ETGs

If the growth results in a sufficient amount of high angular momentum gas
accumulating and cooling in a disc, it may lead to knotty star formation,
Irregular galaxies, and the formation of spiral structures.  This progression
in galaxy evolution is captured by the `Triangal' evolutionary schema for HSB
galaxies \citep{2023MNRAS.522.3588G}, with major `wet' (gas-rich) mergers of
spiral galaxies building bulges and creating massive dust-rich S0 galaxies,
thereby returning spiral galaxies to the ETG classification. Subsequent
major mergers further erase the ordered angular momentum of stellar discs and
disc-like structures to build the elliptical galaxies with their greater
central concentration of stars within the effective half light radius.  ETGs
can also collide with themselves to migrate along the `red sequence', with
major 'dry' (gas-poor) mergers involving supermassive black holes leading to
elliptical galaxies with core-S\'ersic profiles \citep{1980Natur.287..307B,
  2003AJ....125.2951G}.  This behaviour of accretion- and merger-driven galaxy
evolution is apparent in the colour-magnitude diagram
\citep{2024MNRAS.531..230G} for HSB galaxies, including dETGs, and in the
(stellar mass)-(star formation rate) diagram \citep{2024MNRAS.52710059G}.
Such speciation is supported by the distribution of galaxy types in the (black
hole)-bulge mass scaling diagram \citep{2023MNRAS.522.3588G}, which
illustrates how bulges and black holes grow in mass in a monotonic fashion.

In the past, S0 galaxies likely evolved into spiral galaxies when sufficient
fuel was available and able to cool and form stars. Today, S0 galaxies
generally remain S0 galaxies if their H{\footnotesize I} gas disc is either
stable or has been removed.  That is, the bulk of the S0 to spiral
transformations may have already occurred when the Universe was more gas-rich.
While \citet{2009ApJ...699....1L} report the gas-disc in DDO~221, aka WLM
(Wolf-Lundmark-Melotte), has a (rotation speed)-to-(velocity dispersion) ratio
of
$\approx$6.7, compared to $\approx$1.2 for the galaxy's evolved red giant branch
population, the scarcity of dS galaxies reveals that (gas discs built from the
collapse of high angular momentum gas and the ensuing) spiral formation has
been more common in the HSB galaxies, with their higher densities.  This
scenario (see Figure~\ref{Fig12}) does not suppose UDGs to be faded spiral
galaxies, but instead failed spiral galaxies that did not experience
sufficient cold gas accretion and mergers (from, say, satellites or dwarf
spheroidal galaxies) to develop showy discs.

To briefly illustrate this point, two additional samples are used. 
One consists of the UDGs and NUDGes detected in
H{\footnotesize I} and modelled by \citet{2017ApJ...842..133L}.  The 
second sample includes the discs of HSB and LSB spiral (and a few S0) galaxies
from \citet{2001ApJ...556..177G}. 
The scaling parameters for these
relatively gas-rich, 
H{\footnotesize I}-bearing UDGs and NUDGes
have come from \citet{2017ApJ...842..133L}, who fit exponential ($n=1$)
models to their light profiles.  Their $g$-band central surface brightnesses
were corrected for Galactic extinction and also, here, cosmological surface
brightness dimming.  These values were then converted to the $B$-band using the
Galactic extinction corrected $(g-r)_{\rm AB}$ colours from
\citet{2017ApJ...842..133L} and the relation $(B-V)_{\rm Vega} = 0.98(g-r)_{\rm AB} + 0.22$
\citep{2005AJ....130..873J}, along with $(g-B)_{\rm AB} =
-0.366(B-V)_{\rm Vega} - 0.126$ \citep{2006A&A...460..339J} and $B_{\rm
  Vega} = B_{\rm AB}+0.12$.
This conversion gives a $B_{\rm Vega} - g_{\rm AB}$ distribution peaking at
$\approx$0.46 mag, with a FWHM of $\approx$0.06 mag and a slight blue excess
noticed at $\approx$0.33 mag.
From the predominantly spiral galaxy sample, the exponential galaxy discs 
have had their $\mu_{\rm 0,B}$ values and scalelengths, $h_{\rm
  disc}$, corrected following equations~1 and 2 in
\citet{2008MNRAS.388.1708G}, and $h_{\rm disc}$ was converted into $R_{\rm
  e,maj}$ using the multiplier 1.678 \citep[e.g.][their equation
  16]{2005PASA...22..118G}.

For a given size, the 
isolated H{\footnotesize I}-bearing UDGs and NUDGes \citep{2017ApJ...842..133L}
tend to have central surface brightnesses a couple of mag arcsec$^{-2}$
brighter than the average UDG presented thus far \citep[see][their
  figure~7]{2021AnA...654A.105M}.  These H{\footnotesize I}-bearing UDGs
bridge the UDG/NUDG samples from \citet{2024MNRAS.529.3210B} and
\citet{2025MNRAS.536.2536B} and the spiral galaxies in the $\mu_{\rm
  0,B}$-$R_{\rm e}$ diagram (Figure~\ref{Fig12}).
The triangular shaped pathway shown in the lower left of Figure~\ref{Fig12}
reveals how accretion and minor and major mergers can build galaxies.

\section*{Acknowledgements}

This paper is dedicated to Ena (n\'ee McKenzie) Graham (1931-2019), who
adopted AWG long ago and graciously let him claim Scrabble victories during 
her multiple surgeries and lengthy hospital stays from 2017 to 2019.
AWG is grateful to the Department of Astronomy at the University of Florida,
Gainesville, USA, where he largely completed the figures during 2019. 
The delayed write-up has benefitted from an expanded UDG data set and updated
entries in the figures. 
Publication costs were funded through the Australian and New Zealand 
Institutions (Council of Australian University Librarians affiliated) Open   
Access Agreement.
This research has used the NASA/IPAC Extragalactic Database (NED) and
the SAO/NASA Astrophysics Data System (ADS) bibliographic services.

\bibliographystyle{mnras}
\bibliography{Paper-UDG}

\end{document}